\documentclass[twocolumn,prb,showpacs,preprintnumbers,superscriptaddress,amsmath,amssymb,citeautoscript]{revtex4-1}
\usepackage{graphicx}
\usepackage{epstopdf}
\usepackage{dcolumn}
\usepackage{color}
\usepackage{bm}
\usepackage{multirow}
\usepackage{rotating}
\usepackage{array}

\newcommand{\bpm}{\begin{pmatrix}}
\newcommand{\epm}{\end{pmatrix}}
\newcommand{\bs}{\boldsymbol}

\DeclareMathOperator{\sgn}{sgn}
\DeclareMathOperator{\re}{Re}
\DeclareMathOperator{\im}{Im}
\DeclareMathOperator{\ch}{ch}
\DeclareMathOperator{\sh}{sh}

\DeclareMathOperator{\arctg}{arctg}
\DeclareMathOperator{\logn}{ln}
\DeclareMathOperator{\trans}{T}

\begin{document}

\title{Topology from Triviality}
\author{V. Kaladzhyan}
\email{vardan.kaladzhyan@phystech.edu}
\affiliation{Institut de Physique Th\'eorique, Universit\'e Paris Saclay, CEA
CNRS, Orme des Merisiers, 91190 Gif-sur-Yvette Cedex, France}
\affiliation{Laboratoire de Physique des Solides, CNRS, Univ. Paris-Sud, Universit\'e Paris-Saclay, 91405 Orsay Cedex, France}
\affiliation{Department of Physics, KTH Royal Institute of Technology, Stockholm, SE-106 91 Sweden}
\affiliation{Moscow Institute of Physics and Technology, Dolgoprudny 141700, Moscow region, Russia}
\author{Cristina Bena}
\affiliation{Institut de Physique Th\'eorique, Universit\'e Paris Saclay, CEA
CNRS, Orme des Merisiers, 91190 Gif-sur-Yvette Cedex, France}
\author{Pascal Simon}
\affiliation{Laboratoire de Physique des Solides, CNRS, Univ. Paris-Sud, Universit\'e Paris-Saclay, 91405 Orsay Cedex, France}

\date{\today}

\begin{abstract}
We show that bringing into proximity two topologically \emph{trivial} systems can give rise to a \emph{topological} phase. More specifically, we study a 1D metallic nanowire proximitized by a 2D superconducting substrate with a mixed $s$-wave and $p$-wave pairing, and we demonstrate both analytically and numerically 
that the phase diagram of such a setup can be richer than reported before. Thus, apart from the two ''expected'' well-known phases (i.e., where the substrate and the wire are both simultaneously trivial or topological), we show that there exist two peculiar phases in which  the nanowire can be in a topological regime while the substrate is trivial, and vice versa. 
\end{abstract}

\maketitle

\section{Introduction} The last decade was marked by numerous proposals to realize time-reversal-symmetric (TRS) topological superconductors \cite{Qi2009,Qi2010,Deng2012,Wong2012,Zhang2013,
Keselman2013,Chung2013,Liu2014,Gaidamauskas2014,Dumitrescu2014,
Haim2014,Klinovaja2014,Klinovaja2014-2} hosting so-called Majorana Kramers pairs. Despite the fact that all the braiding operations in such systems are bound to be performed with \emph{pairs} rather than with single Majorana quasiparticles, it has been proposed theoretically that braiding of end states remains non-Abelian due to the protection by time reversal symmetry \cite{Liu2014}. 

Certain recipes for creating TRS topological superconductivity (SC) are of high relevance to the current manuscript. First, it is known that a metallic nanowire (NW) proximitized by a SC becomes superconducting due to the proximity effect. Moreover, when the substrate is topological, Majorana fermions may form in the wire  \cite{Chevallier2013}. Furthermore, as shown in Ref.~[\onlinecite{Nakosai2013prl}], a NW proximitized by a helical $p$-wave SC substrate becomes a 1D TRS topological SC exhibiting triplet pairing and Majorana Kramers pairs at its ends. It was also demonstrated in \cite{Liu2014} that if the SC substrate is characterized by a mixed singlet-triplet pairing with a dominant triplet component, then the NW becomes topological.

Second, in a recent paper, Neupert \emph{et al.} considered a chain of scalar impurities immersed into a superconductor with a mixed $s$-wave and $p$-wave order parameter \cite{Neupert2016}. Scalar impurities do not give rise to Yu-Shiba-Rusinov states \cite{Yu1965,Shiba1968,Rusinov1969} in purely $s$-wave SCs, in accordance with the Anderson theorem \cite{Anderson1959}. However, in the presence of a $p$-wave pairing component, a pair of in-gap bound states is formed in the presence of a scalar impurity\cite{Balatsky2006,Kaladzhyan2015,Kaladzhyan2016a}. Given a chain of such impurities, the corresponding bound states hybridize into a so-called "Shiba band". As shown in Ref.~[\onlinecite{Neupert2016}], 
such a band can enter a topological regime, supporting Majorana bound states at its ends, even though the superconductor is in a trivial phase with a dominant $s$-wave component.

Furthermore, Hsieh \emph{et al.} demonstrated that topological phases can be induced in trivial systems via coupling to topological ones, despite the absence of a local order parameter in the latter. One of the examples in their paper \cite{Hsieh2016} focuses on how a trivial system coupled to a Chern insulator in a topological regime with Chern number $-1$ becomes topological with an opposite Chern number, $+1$. The authors also proposed an analogous example with two coupled 1D systems instead of 2D ones.

\begin{figure}
	\includegraphics[width=0.95\columnwidth]{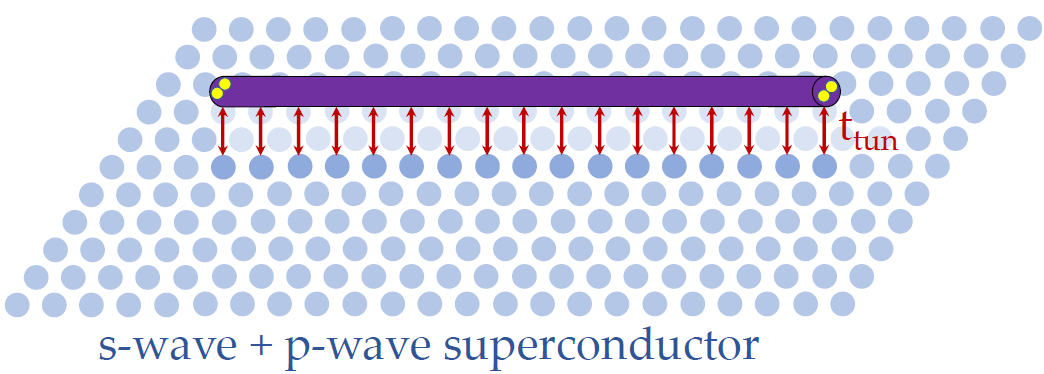}
	\caption{(Color online) A sketch of the system: a metallic nanowire (violet) on top of a 2D SC substrate with mixed singlet and triplet pairing (blue). Red arrows denote the tunnelling between the substrate and the wire, and the Majorana Kramers pairs are drawn as yellow spheres at the ends of the wire.}
	\label{systemsketch}
\end{figure}

Motivated by the aforementioned findings, we offer in this paper another missing piece of the puzzle. In what follows we show that bringing into proximity two topologically \emph{trivial} systems can give rise to a \emph{topological} phase.  We exemplify this idea by studying a 1D metallic NW proximitized by a 2D SC substrate with a mixed $s$-wave and $p$-wave pairing (see Fig.~\ref{systemsketch}).
We obtain the phase diagram of the system in Fig.~\ref{systemsketch} using both analytical and numerical methods. The analytical approach is based on integrating out the SC degrees of freedom and obtaining an effective low-energy Hamiltonian for the nanowire, whereas the numerical simulations are done within a tight-binding model on a square lattice. The phase diagrams are constructed by testing the formation of Majorana bound states at the ends of the nanowire.

The resulting phase diagrams recover first the two phases that have also been previously reported  \cite{Nakosai2013prl,Liu2014}: one in which both the substrate and the nanowire are trivial, as well as one in which both are topological. However, they exhibit also two salient unexpected phases which have never been discussed before. Thus we identify a phase in which the SC is topological, whereas the induced superconductivity in the NW is trivial. Also, most surprisingly, one can achieve a phase in which the SC substrate is trivial whereas the NW is topological.  \\

\section{Continuum model} We start by considering a 1D metallic NW  oriented along the $x$ axis and  deposited on top of a 2D TRS SC with both $s$-wave and $p$-wave pairing, lying in the $(x,y)$ plane. The  Hamiltonian of the NW is given by
\begin{equation}
\mathcal{H}_{\mathrm{NW}} = \left(\frac{p_x^2}{2m} - \mu\right) \sigma_0 \otimes \tau_z ,
\label{HNW}
\end{equation}
and the Hamiltonian of the superconducting substrate can be written as
\begin{equation}
\mathcal{H}_{\mathrm{SC}} = \frac{\bs{p}^2-p_F^2}{2M}\sigma_0 \otimes \tau_z + \Delta_s \sigma_0 \otimes \tau_x + \varkappa (p_y \sigma_x - p_x \sigma_y) \otimes \tau_x .
\label{HSC}
\end{equation}
We use the Nambu bases $\left\{ c_{p_x \uparrow },c_{ p_x \downarrow},c^{\dag}_{-p_x \downarrow},-c^{\dag}_{-p_x \uparrow} \right\}^{\trans}$ where $c_{p_x \uparrow }$ destroys an electron of spin $\uparrow$ with momentum $p_x$ in the NW, and  $\left\{ \Psi_{\bs{p} \uparrow },\Psi_{\bs{p} \downarrow },\Psi^{\dag}_{- \bs{p}\downarrow },-\Psi^{\dag}_{- \bs{p} \uparrow } \right\}^{\trans}$ for which $ \Psi_{\bs{p} \uparrow }$ destroys an electron of momentum $\bs{p}$ in the substrate.
The matrices ${\bs\sigma}$ and ${\bs\tau}$ are the Pauli matrices acting respectively in the spin and the particle-hole subspaces. 
In \eqref{HNW}, 
$m$ is the effective mass of the quasi-particles in the NW, and $\mu$ is their chemical potential. Similarly, in \eqref{HSC}, $M$ and $p_F$ embody the effective mass  of quasi-particles in the SC and their Fermi momentum. The singlet and triplet SC pairings in the substrate are denoted by $\Delta_s$ and $\varkappa$, respectively. We model the coupling between the NW and the SC by means of a tunnelling Hamiltonian
\begin{equation}
\mathcal{H}_{\trans} = t_{\mathrm{tun}}\sum\limits_{\bs{p} \sigma} \left[ \Psi^\dag_{\bs p \sigma} c_{p_x\sigma} + c^\dag_{p_x\sigma} \Psi_{\bs p \sigma} \right],
\label{HTun}
\end{equation} 
where $\sigma = \uparrow,\downarrow$, and $t_{\mathrm{tun}}$ is the tunnelling amplitude.\\

In what follows we integrate out the SC degrees of freedom (see Appendix \ref{AppendixDerivation}) and we obtain an effective retarded Green's function for the proximitized NW:
\begin{equation}
\left[G^{\mathrm{eff}}_{\mathrm{NW}}\left(\omega, p_x \right)\right]^{-1} = \left( G_{\mathrm{NW}} \right)^{-1} - \int\negthickspace \frac{dp_y}{2\pi} \mathcal{H}_\mathrm{T}^\dag G_{\mathrm{SC}} \mathcal{H}_\mathrm{T}
\label{GeffNW}
\end{equation}
where we defined 
$
G_{\mathrm{NW}} \equiv \left(\omega + i0 - \mathcal{H}_{\mathrm{NW}} \right)^{-1}$, and  $G_{\mathrm{SC}} \equiv \left(\omega + i0 - \mathcal{H}_{\mathrm{SC}} \right)^{-1}.
$
The effective Green's function given by Eq.~(\ref{GeffNW}) yields an effective low-energy Hamiltonian describing the NW (see, e.g., Ref.~\cite{Nakosai2013prl}):
\begin{equation}
\mathcal{H}^{\mathrm{eff}}_{\mathrm{NW}} \equiv - \left(G^{\mathrm{eff}}_{\mathrm{NW}}\right)^{-1} \left|_{\omega = 0} \right.,
\label{HeffNWdef}
\end{equation}
where we keep all the terms up to those linear in $p_x$. It is worth mentioning that setting the frequency to zero in Eq.~(\ref{HeffNWdef}) is an approximation that allows us to extract a qualitative effective Hamiltonian for the NW. More accurate results can be obtained from an exact diagonalization of the full lattice model, however, as we will show in what follows, the two methods yield consistent results.\\

{\it Results.} We leave the detailed presentation of the tedious calculations for Appendix \ref{AppendixDerivation}, and we present here the first result of this paper -- an effective low-energy Hamiltonian for the NW:
\begin{align}
\nonumber H_{\mathrm{NW}}^{\mathrm{eff}} = \left( \frac{p_x^2}{2m}-\mu^{\mathrm{ind}} \right) \sigma_0 \otimes \tau_z - \lambda^{\mathrm{ind}} \, p_x \sigma_y \otimes \tau_z + \\
 + \Delta_s^{\mathrm{ind}} \, \sigma_0 \otimes \tau_x - \varkappa^{\mathrm{ind}} \,  p_x \sigma_y \otimes \tau_x. \phantom{aaa}
\label{HeffNW}
\end{align}
As a result of the interplay between the $s$-wave and $p$-wave pairing,  an effective Rashba-like spin-orbit coupling arises in the NW. We note that for a pure $p$-wave or pure $s$-wave coupling no such term is generated, provided there is no spin-orbit coupling in the substrate. The induced parameters in Eq.~(\ref{HeffNW}) are defined as follows:
\begin{eqnarray}
\Delta_s^{\mathrm{ind}}& \equiv& \frac{t_{\mathrm{tun}}^2}{\pi v_F} \sum\limits_{\sigma = \pm} \re \left[ \frac{\sgn\Delta_\sigma - i\sigma \tilde{\varkappa}}{1+\tilde{\varkappa}^2}\arctg \frac{-|p_\varepsilon| + z_\sigma}{\sqrt{p_\varepsilon^2-z_\sigma^2}}\right] \nonumber\\
\varkappa^{\mathrm{ind}} &\equiv& \frac{t_{\mathrm{tun}}^2}{\pi v_F}\sum\limits_{\sigma = \pm} \re \left[ \frac{-\sigma\sgn\Delta_\sigma + i\tilde{\varkappa}}{1+\tilde{\varkappa}^2} \frac{1}{z_\sigma}\arctg \frac{-|p_\varepsilon| + z_\sigma}{\sqrt{p_\varepsilon^2-z_\sigma^2}}\right] \nonumber\\
\lambda^{\mathrm{ind}} &\equiv& \frac{t_{\mathrm{tun}}^2}{\pi v_F}  \sum\limits_{\sigma = \pm}  \im \left[\frac{-\sigma + i \tilde{\varkappa} \sgn\Delta_\sigma}{1+\tilde{\varkappa}^2} \frac{1}{z_\sigma}\arctg \frac{-|p_\varepsilon| + z_\sigma}{\sqrt{p_\varepsilon^2-z_\sigma^2}}\right] \nonumber\\
\nonumber \mu^{\mathrm{ind}} &\equiv &\frac{t_{\mathrm{tun}}^2}{\pi v_F} \sum\limits_{\sigma = \pm} \im \left[\frac{i\sigma \tilde{\varkappa} \sgn\Delta_\sigma-1}{1+\tilde{\varkappa}^2}\arctg \frac{-|p_\varepsilon| + z_\sigma}{\sqrt{p_\varepsilon^2-z_\sigma^2}} \right] + \nonumber\\
&&\phantom{aaaaaaaaa}+ \mu + \frac{t_{\mathrm{tun}}^2}{\pi v_F} \frac{\logn 2\Lambda}{1+\tilde{\varkappa}^2},
\label{ind}
\end{eqnarray}
where $\Delta_\sigma \equiv \Delta_s + \sigma \varkappa p_F$, $\tilde{\varkappa} \equiv \varkappa/v_F$, while $p_\varepsilon$ and $\Lambda$ are the IR and UV momentum cutoffs, respectively, unavoidable in an effective low-energy theory. Finally, we have
$
z_\sigma =[\sigma \varkappa \Delta_s - p_F v_F^2 + i v_F |\Delta_\sigma|][v_F^2 + \varkappa^2]^{-1}.
$
We note that the induced chemical potential depends weakly on the UV momentum cutoff $\Lambda$, 
however, this term can be absorbed into a redefinition of the initial chemical potential $\mu$ of the NW. 
Our results are fully consistent with calculations performed in Ref.~[\onlinecite{Nakosai2013prl}] in the limit $\Delta_s = 0$, namely when the substrate is a purely triplet SC.\\

To find the induced SC gap for the Hamiltonian in Eq.~(\ref{HeffNW}) we linearize its spectrum around the two different Fermi momenta emerging due to the non-zero induced spin-orbit coupling. We thus obtain\footnote{Even though this value is close to the actual value of the induced gap, we should note that this expression is however an approximation valid only when the spectrum can be linearized.}:
\begin{align}
\Delta^{\mathrm{ind}}_{\mathrm{eff}} = \min\limits_{\sigma = \pm}\frac{ \left| \Delta^{\mathrm{ind}}_s - m \varkappa^{\mathrm{ind}} \left(\lambda^{\mathrm{ind}} - \sigma \sqrt{(\lambda^{\mathrm{ind}})^2 + \frac{2\mu^{\mathrm{ind}}}{m}}\right) \right|}{\sqrt{1+\frac{(\varkappa^{\mathrm{ind}})^2}{(\lambda^{\mathrm{ind}})^2 + \frac{2\mu^{\mathrm{ind}}}{m}}}}.
\end{align}

To test the formation of a topological phase in the NW we check first that the induced effective gap value $\Delta^{\mathrm{ind}}_{\mathrm{eff}}$ is non-zero, and second, that the Hamiltonian (\ref{HeffNW}) has a localized zero-energy solution. The underlying superconductor is topological when $\Delta_s < \varkappa p_F$ \cite{Sato2009} and the NW if $ |\Delta^{\mathrm{ind}}_s|< |\varkappa^{\mathrm{ind}} p_{\sigma_0}^{\mathrm{ind}}|$ with   $p_{\sigma_0}^{\mathrm{ind}} = m \left(\lambda^{\mathrm{ind}} - \sigma_0 \sqrt{(\lambda^{\mathrm{ind}})^2 + \frac{2\mu^{\mathrm{ind}}}{m}}\right)$ is the corresponding effective Fermi momentum in the wire, obtained by taking into account also non-zero Rashba spin-orbit coupling $\lambda^{\mathrm{ind}}$ and the corresponding split in the band structure. $\sigma_0$ is the value of $\sigma$ that minimizes the induced gap $\Delta^{\mathrm{ind}}_{\mathrm{eff}}$.  

As we will show below, the most important factor in obtaining different topological phases in  the wire and in the substrate is the difference between  $|p_{\sigma_0}^{\mathrm{ind}}|$ and $p_F$. The former depends only weakly on $p_F$ (see Fig.~\ref{figureA1} in Appendix \ref{AppendixAnalysis}), but very strongly on $\mu$, which can thus be used as a "topological knob".  

The induced chemical potential of the NW $\mu^{\mathrm{ind}}$ is also affected by the substrate, and depends linearly on the initial chemical potential of the NW $\mu$ (see Eq.~(\ref{ind})).  

\begin{figure}
	\centering		
	\includegraphics[width=0.9\columnwidth]{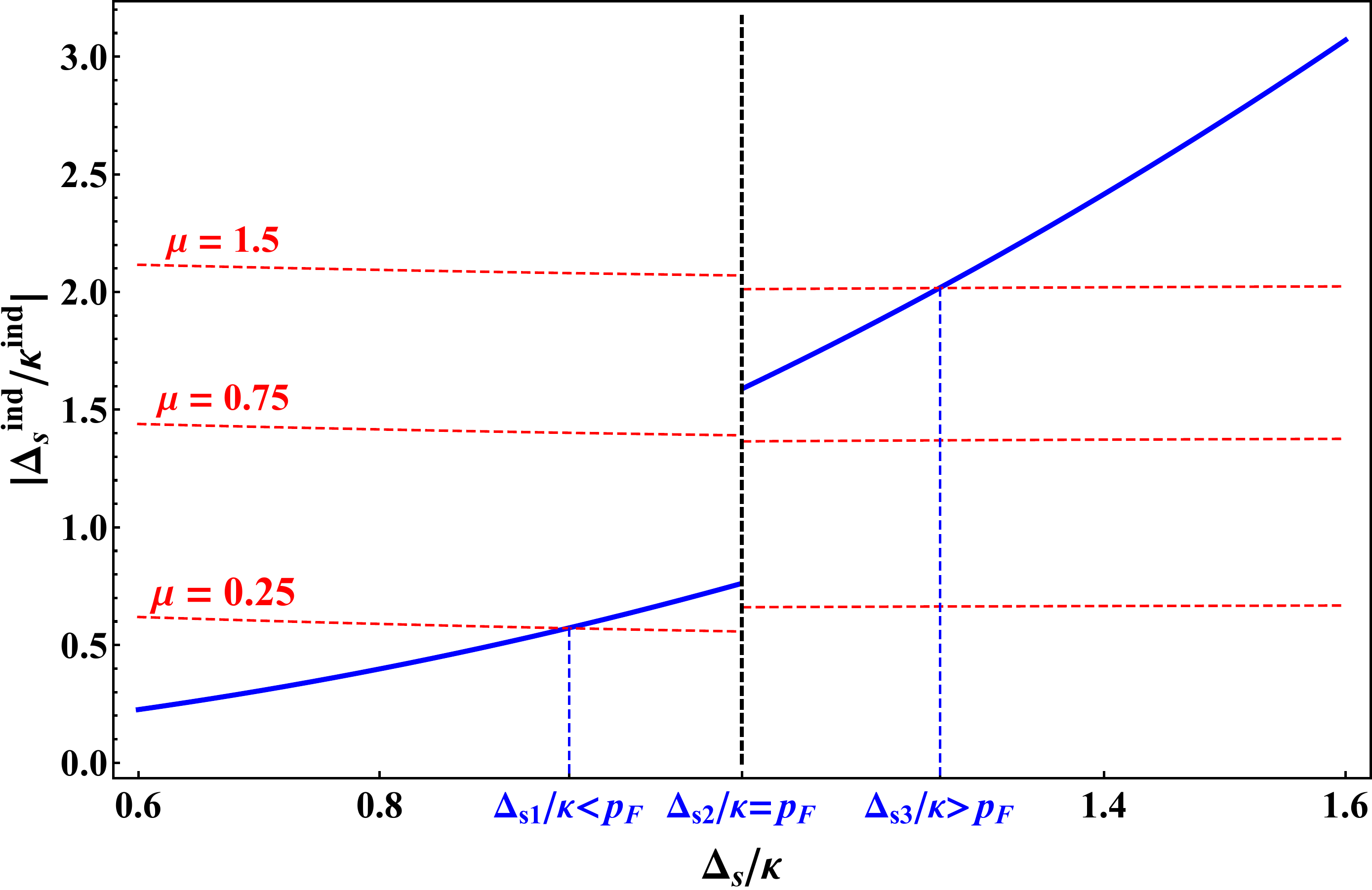}
	\caption{(Color online) The ratio of induced pairing parameters $|\Delta_s^{\mathrm{ind}}/\varkappa^{\mathrm{ind}}|$, as a function of $\Delta_s/\varkappa$ (blue lines). The black dashed vertical line marks the topological phase transition in the substrate. The red dashed lines describe the dependence of $|p_{\sigma_0}^{\mathrm{ind}}|$ on $\Delta_s$  for three different values of the chemical potential in the NW. For each value of $\mu$ we have indicated the intersection points corresponding to the values of the critical $\Delta_s$ in the substrate which would yield a topological phase transition in the NW. Note that the critical $\Delta_s$ of the NW can be smaller, larger or equal to that of the substrate. We have set $M=1.5, \varkappa=2.5, p_F=1.1$, and $t_{\mathrm{tun}}=0.5$.}
	\label{ratioplot}
\end{figure}

To understand the mechanisms behind the induced phase transition, in Fig.~\ref{ratioplot}, we plot $|\Delta_s^{\mathrm{ind}}| / |\varkappa^{\mathrm{ind}}|$ as a function of $\Delta_s$ in the substrate, normalized in units of $\varkappa$ in the substrate. The substrate topological phase transition takes place at $\Delta_s/\varkappa=p_F$, the Fermi momentum of the substrate, marked by the vertical black dashed line in the plot. Note the interesting discontinuity at $\Delta_s/\varkappa=p_F$, we are going to discuss it in more detail in what follows. 

The critical $\Delta_s$ in the substrate corresponding to the topological phase transition in the NW can be determined by finding the points on this graph for which $|\Delta_s^{\mathrm{ind}}| / |\varkappa^{\mathrm{ind}}|= |p_{\sigma_0}^{\mathrm{ind}}|$. The dependence of $|p_{\sigma_0}^{\mathrm{ind}}|$ on $\Delta_s$ in the substrate is depicted in Fig.~\ref{ratioplot} by the dashed red lines corresponding to three different chemical potentials in the NW. Note that $|p_{\sigma_0}^{\mathrm{ind}}|$ is almost independent of $\Delta_s$, except for small jump at $\varkappa p_F$, its main variation is with $\mu$. The critical values at which the NW becomes topological are obtained by taking the intersection points of the two graphs, $|\Delta_s^{\mathrm{ind}}| / |\varkappa^{\mathrm{ind}}|$ and $|p_{\sigma_0}^{\mathrm{ind}}|$ (see Fig.~\ref{ratioplot}). We see that by varying $\mu$ in the NW we can tune the intersection points and thus the critical $\Delta_s$ for which the NW becomes topological and make it smaller or larger than the one in the substrate.  A very interesting point to make is that for values of $\mu$ yielding an intersection point inside the discontinuity, the two critical $\Delta_s$'s, for the NW and the substrate, are identical.

Thus, we can see that by tuning $\mu$ in the NW we can tune the transition such that we obtain a topological phase in the NW, associated with a trivial phase in the SC. From the above discussion we can conclude that obtaining such a phase is quite a generic feature as long as, in the first place, one can tune $p_{\sigma_0}^{\mathrm{ind}}$ independently of $p_F$. 
Moreover, $p_{\sigma_0}^{\mathrm{ind}}$ depends quite strongly on the chemical potential in the NW, and for our chosen parameters we can approximate $|p_{\sigma_0}^{\mathrm{ind}}| \propto \sqrt{\mu^{\mathrm{ind}}} \propto \sqrt{\mu}$. Thus, except for the very peculiar and most likely improbable situation in which $|\Delta_s^{\mathrm{ind}}| / |\varkappa^{\mathrm{ind}}|=\Delta_s/\varkappa$, and $p_{\sigma_0}^{\mathrm{ind}}=p_F$, one would be able to generate the unusual phases with different topological properties for the NW and the substrate. 

As mentioned above, the second necessary ingredient to obtain different topological phases in the NW and the substrate is that the discontinuity in the $|\Delta_s^{\mathrm{ind}}| / |\varkappa^{\mathrm{ind}}|$ does not cover the entire range of achievable physical values for $p_{\sigma_0}^{\mathrm{ind}}$. Unfortunately it is hard to predict its exact dependence on the parameters of the system analytically, but a simple numerical analysis using Eq.~(\ref{ind}) from the main text allows to extract its asymptotic behavior. We have found that the absolute value of the jump is almost exactly given by $2/\varkappa$, with the larger $\varkappa$, the more precise the estimation. Thus increasing the triplet pairing in the substrate may also increase the range in which the NW is topological while the substrate is not.

An interesting note to make is that, by examining Fig.~\ref{ratioplot}, we can see that when the substrate is in a topological regime (i.e. on the left side of the dashed line), none of the pairings in the NW is favored (mathematically speaking, $|\Delta_s^{\mathrm{ind}}/\varkappa^{\mathrm{ind}}| \approx \Delta_s/\varkappa$). However, in the trivial regime (i.e. on the right side of the dashed line) we can infer from the slope of the curve that inducing singlet pairing is more favorable than triplet one, and thus we would intuitively expect a trivial phase in the wire in this range. However, as it turns out, $p_{\sigma_0}^{\mathrm{ind}}$, and not $|\Delta_s^{\mathrm{ind}}/\varkappa^{\mathrm{ind}}|$ is the most important knob for tuning the phase transition of the NW, and it is its variation that allows us to obtain the non-trivial combination of phases discussed above.

In what follows we construct a phase diagram capturing the topological character of both the SC and the NW. We emphasise that we do not study topological properties of the system "substrate + nanowire" as a whole, but we study them separately, showing both simultaneously on the phase diagrams. This can be achieved by combining the $\mathbb{Z}_2$ topological index for the NW and the $\mathbb{Z}_2$ index for the substrate into a global topological index $\delta$ as follows. The topological character of each system is considered separately, by means of a standard criterion mentioned above, i.e. the sign of $\Delta_s - \kappa p_F$ shows the nature of a given phase. We combine two criteria into one by summing them up and introducing factors $3/2$ for the substrate and $1/2$ for the wire, solely for convenience, to distinguish the case of topological substrate and trivial nanowire from that of trivial substrate and topological nanowire (otherwise the sum of topological indices would give $0$ in both cases). This combined index $\delta$ allows to discriminate between all four different combinations of phases we can possibly expect in our setup, and is thus given by:
\begin{equation}
\delta = \frac{3}{2} \sgn \left(\Delta_s - \varkappa p_F \right) + \frac{1}{2} \sgn \left(|\varkappa^{\mathrm{ind}} p_{\sigma_0}^{\mathrm{ind}}| - |\Delta^{\mathrm{ind}}_s| \right).
\label{topindexdef}
\end{equation}
$\delta$ can take four different values, corresponding to the four possible phases of the NW-SC systems; these values and the corresponding phases are summarized in Table I. The most exciting phase  corresponds to $\delta=+2$, for which a NW can attain a topological phase when proximitized by a trivial SC.
\begin{center}
\begin{table}
	\begin{tabular}{|llll|llll|l|}
	\hline
	            &&&& SC  &&&& NW \\ \hline
	$\delta=+2$ &&&& trivial &&&& topological \\ \hline
	$\delta=+1$ &&&& trivial &&&& trivial \\\hline
	$\delta=-1$ &&&& topological &&&& topological \\ \hline
	$\delta=-2$ &&&& topological &&&& trivial \\\hline
	\end{tabular} 
	\caption{The four possible topological phases for the NW-SC system.}
	\label{table}
	\end{table}
\end{center}

\begin{figure}
	\centering
	\includegraphics[width=0.7\columnwidth]{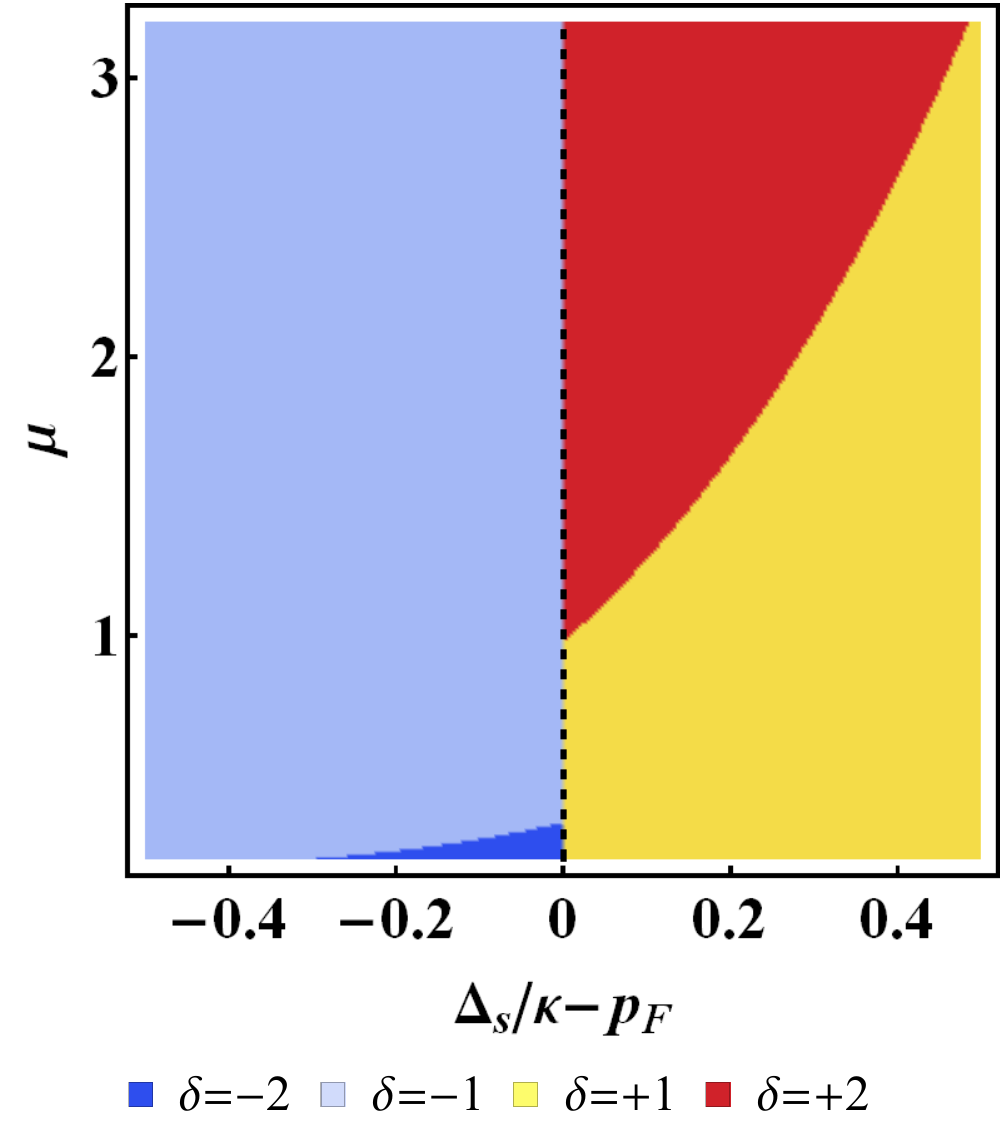}
	\caption{(Color online) The phase diagram of a 1D metallic NW deposited on top of a 2D SC substrate. We plot the value of the topological index $\delta$ defined in Eq.~(\ref{topindexdef}) as a function of  $\frac{\Delta_s}{\varkappa}-p_F$ in the substrate, and of $\mu$ in the NW. We note that all the four possible phases corresponding to $\delta = \pm 1, \pm 2$ can form. The region with $\delta=+2$ (denoted in red) is the most interesting, and corresponds to a {\it topological} phase of the NW induced via the proximity of a {\it trivial}  SC. The black dashed line marks the bulk topological phase transition for the substrate. We have set $p_F=1.1,  M=m=1.5, \varkappa=2.5$, and $t_{\mathrm{tun}}=0.5$.}
	\label{phdanalyt}
\end{figure}

In Fig.~\ref{phdanalyt} we plot the value of the topological index $\delta$ as a function of $\Delta_s$ (more precisely as a function of $\frac{\Delta_s}{\varkappa}-p_F$), and of $\mu$, when all the other parameters of the model are fixed. We note the formation of all the four possible phases that could be realized in this setup. First two combinations were known and studied before: "trivial substrate + trivial nanowire" (the yellow phase with $\delta = +1$ in Fig.~\ref{phdanalyt}) and "topological substrate + topological nanowire" (the light blue phase with $\delta = -1$ in Fig.~\ref{phdanalyt}). The main observable signature of the latter combination is the local density of states due to the formation of Majorana zero modes at the ends of the nanowire and at the boundary of the substrate. The other two combinations that we discovered and described in our paper correspond to having "topological substrate + trivial nanowire" (deep blue phase with $\delta = -2$ in Fig.~\ref{phdanalyt}) and "trivial substrate + topological nanowire" (red phase with $\delta = +2$ in Fig.~\ref{phdanalyt}). In the former case Majorana bound states will appear solely at the boundary of the substrate, whereas in the latter combination they will form only at the ends of the nanowire. For any set of parameters corresponding to this phase, we can demonstrate that there is a doubly-degenerate localized zero-energy solution, confirming the topological character of the NW. Due to the presence of TRS these Majorana bound states form protected Kramers pairs which are robust to non-magnetic disorder.

\begin{figure}[h!]
	\centering
	\includegraphics[width=0.72\columnwidth]{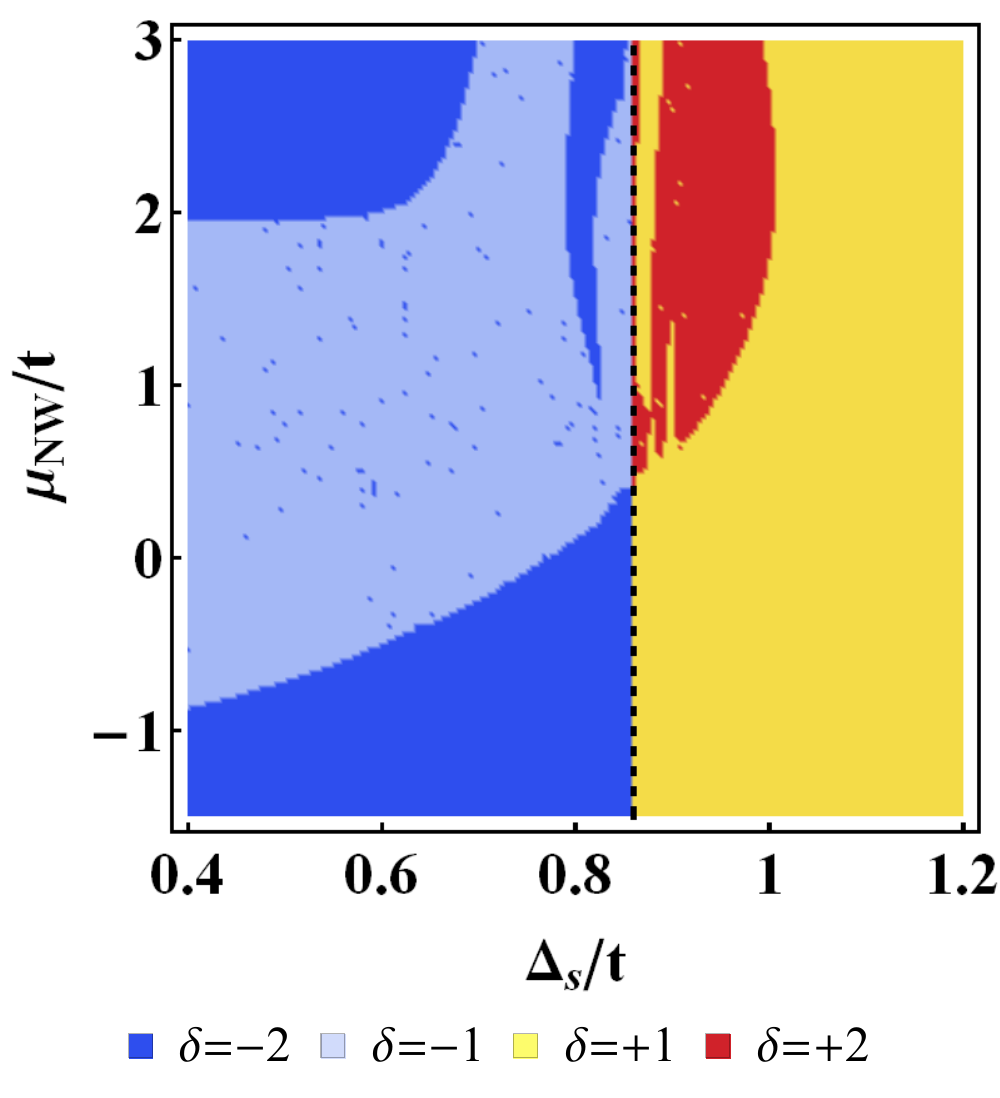}
	\caption{(Color online) The phase diagram of a 1D NW of length $L=151$ deposited on top of a $201 \times 51$  2D SC substrate. We plot the combined topological index $\delta$, as defined in Table I, as a function of $\Delta_s$ and  $\mu_{\mathrm{NW}}$ (in units of $t$, which we set to $1$). The topological character of the NW is obtained by calculating the MP  of the lowest-energy states (with a cutoff at 0.9). The dashed line at $\Delta_s \sim 0.86t$ marks the bulk topological phase transition.  We set $\mu_{\mathrm{SC}}=3, \Delta_t = 0.5$  and $t_{\mathrm{tun}}=2$ .
	}
	\label{phdlattice}
\end{figure}

\section{Lattice model} Hereinafter, we consider a tight-binding model for the system described above. Thus we model the SC substrate as a square lattice of size $W_x \times W_y$, and the NW as a wire of length $L$ lying on top of it, and having the same lattice constant. The electrons can tunnel between each site of the NW and the corresponding site in the SC substrate (see Fig.~1). The tight-binding Hamiltonian is given by
\begin{align}
\nonumber\mathcal{H}_{\mathrm{TB}} = &\sum\limits_{\bs{r}\in SC,\sigma} -\mu_{\mathrm{SC}} \Psi^\dag_{\bs{r}\sigma} \Psi_{\bs{r}\sigma} + \sigma \Delta_s \Psi_{\bs{r}\sigma} \Psi_{\bs{r},-\sigma}  \\
\nonumber &- t\sum\limits_{\bs{r}\in SC,\sigma}\left( \Psi^\dag_{\bs{r}\sigma} \Psi_{\bs{r}+\bs{x},\sigma} +\Psi^\dag_{\bs{r}\sigma} \Psi_{\bs{r}+\bs{y},\sigma} \right)  \\
\nonumber &+\Delta_t \sum\limits_{\bs{r}\in SC,\sigma} \Psi_{\bs{r}\sigma} \Psi_{\bs{r}+\bs{x},\sigma} + i\sigma \Psi_{\bs{r}\sigma} \Psi_{\bs{r}+\bs{y},\sigma}   \\
\nonumber &+\sum\limits_{x\in NW,\sigma} -\mu_{\mathrm{NW}} c^\dag_{x\sigma} c_{x\sigma} - tc^\dag_{x\sigma} c_{x+1,\sigma}   \\
&+ t_{\mathrm{tun}}\sum\limits_{x \in NW,\sigma} \Psi^\dag_{x, y=0,\sigma} c_{x,\sigma} + \mathrm{H.c.}
\label{TBHam}
\end{align}
Here $\Delta_s$ and $\Delta_t$ stand for the singlet and triplet pairing amplitudes in the substrate respectively, $\sigma=\uparrow,\downarrow$, $\bs x$ and $\bs y$ are the unit vectors of the 2D SC lattice, and the lattice constant $a$ is set to unity. The chemical potentials in the substrate and in the wire are denoted as $\mu_{\mathrm{SC}}$ and $\mu_{\mathrm{NW}}$, $t$ is the hopping parameter assumed to be the same in the SC and NW, and $t_{\mathrm{tun}}$ is the tunnelling amplitude between the SC and the NW.

We explore the phase diagram of this system by performing a numerical diagonalization using the MatQ code \cite{matq} (see Fig.~\ref{phdlattice}). Same as before, we plot a combined topological index describing the global character of the system (see Table I). We test the topological character of the NW and the formation of Majorana fermions in the NW using the Majorana polarization (MP) \cite{Sticlet2012,Sedlmayr2015b,Sedlmayr2016,Kaladzhyan2017a,Kaladzhyan2017b}. The topological character of the bulk SC is also calculated numerically and the corresponding bulk phase transition is indicated by a vertical line at $\Delta_s \sim 0.86t$. First we note that the resulting phase diagram agrees qualitatively with the one obtained using the continuum model in the previous section (see Fig.~\ref{phdanalyt}). As before, the sought-for interesting phase is the one for which the NW is topological, whereas the substrate is trivial ($\delta=2$). Such a phase  can be identified here also (depicted in red). Moreover, we recover a region in which the NW is trivial while the bulk SC is topological ($\delta=-2$, depicted in blue).
	 
	It is also important to mention that there are upper and lower boundary values of the chemical potential limiting the existence of the topological regions (such as $\mu=2$ for $\Delta_s=0$ in Fig.~\ref{phdlattice}). Such boundaries are the natural consequence of the finite bandwidth of the lattice model, and cannot be present for the continuum model studied above. 
	
Note that numerical errors and a significant amount of noise may occur close to the bulk phase transition line at $\Delta_s \sim 0.86t$ due to the fact that the SC gap value becomes very small, and therefore the localization length of Majorana modes becomes comparable to the size of the wire $L$ in this regime.

\section{The effects of spin-orbit coupling in the substrate}

\begin{figure}[h!]
	\centering		
	\includegraphics[width=0.72\columnwidth]{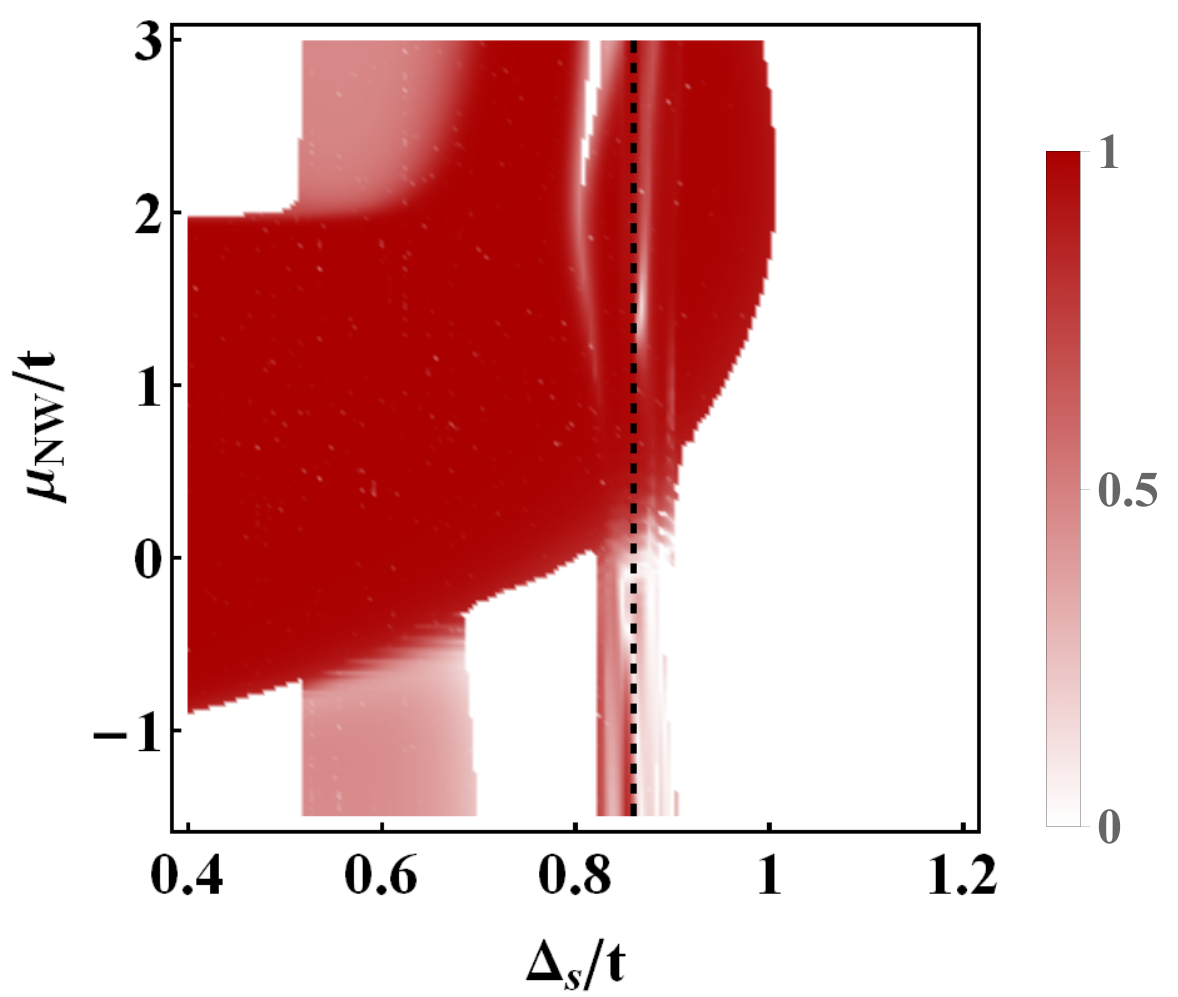}\\	
	\includegraphics[width=0.72\columnwidth]{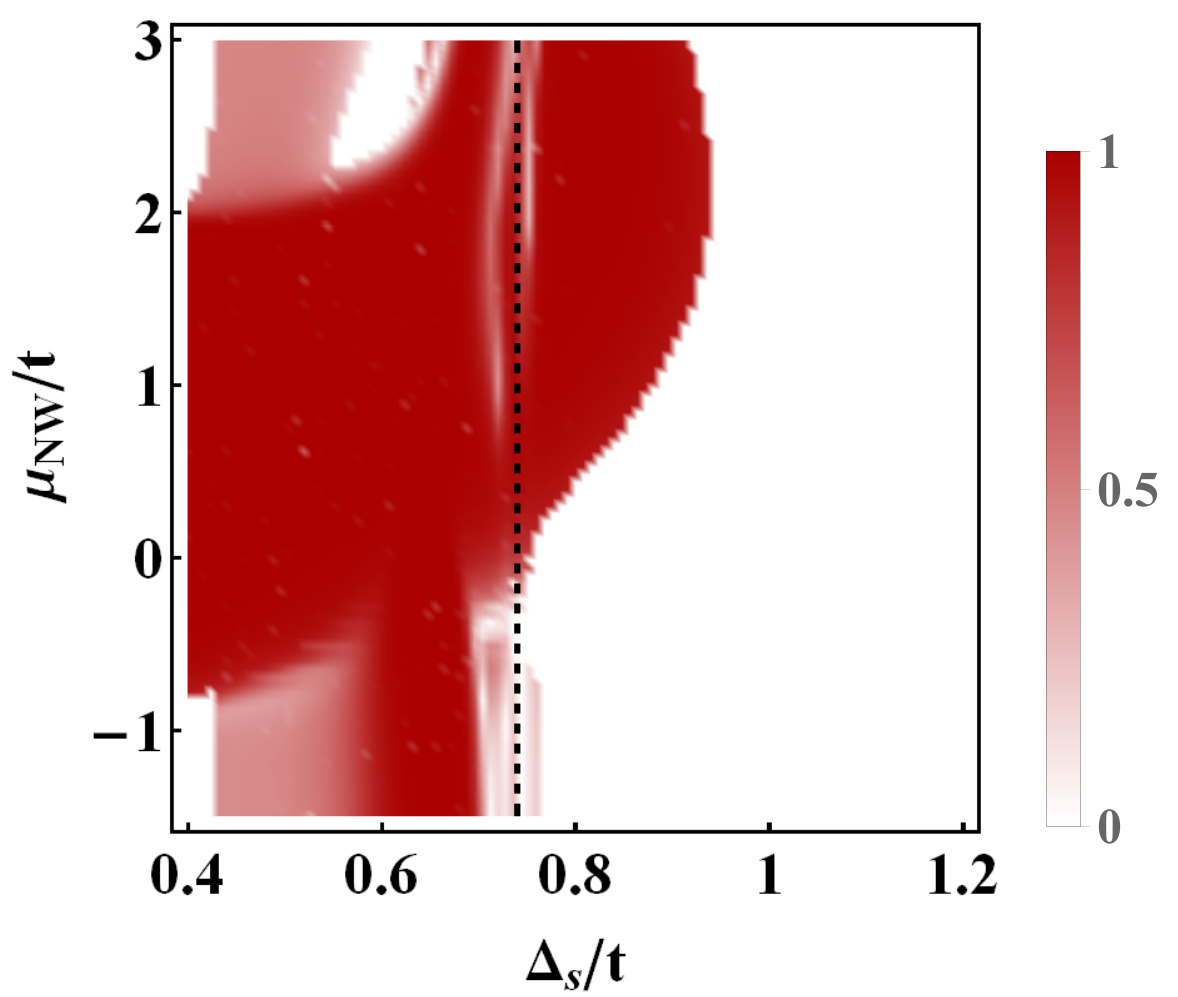}\\
	\includegraphics[width=0.72\columnwidth]{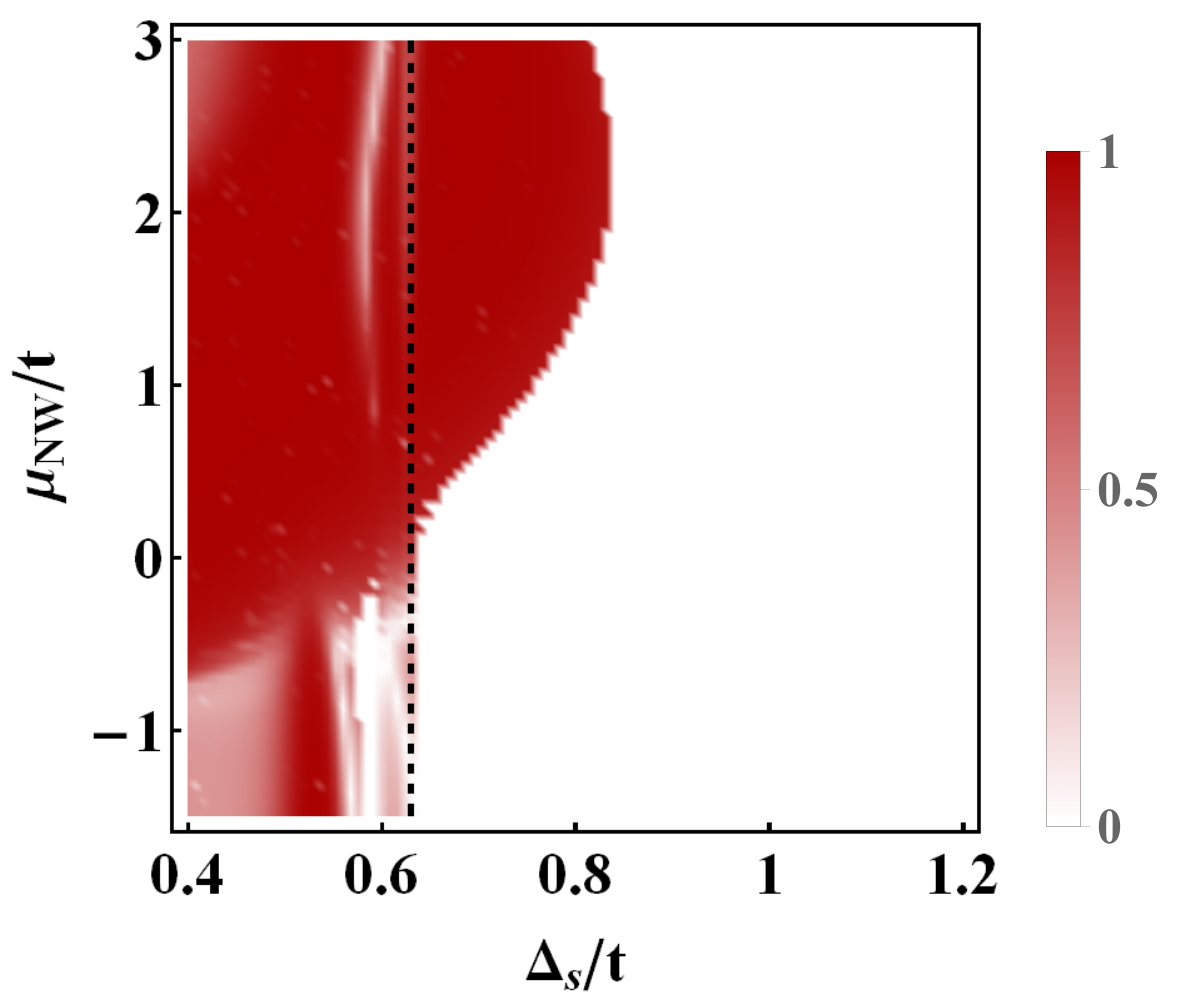}
	\caption{(Color online) The phase diagrams of a 1D metallic NW deposited on top of a 2D SC substrate with a finite Rashba SO coupling ($\lambda=0.2$ and $\lambda=0.4$ in the middle and lower panels respectively) and in its absence (the upper panel). We plot the MP as a function of  $\Delta_s$ and  the chemical potential $\mu_{\mathrm{NW}}$ of the NW (in units of the hopping parameter $t$ which we set to $1$ in our simulations). A MP value of $1$ corresponds to the formation of Majorana Kramers pairs at the ends of the wire, and $0$ to a trivial phase.  The black dashed lines at $\Delta_s \sim 0.86t$ (upper), $\Delta_s \sim 0.74t$ (middle) and  $\Delta_s \sim 0.63t$ (lower) mark the bulk topological phase transition.  We set $\mu_{\mathrm{SC}}=3, \Delta_t = 0.5$  and $t_{\mathrm{tun}}=2$.}
	\label{phdSO}
\end{figure}

Another important aspect of our proposal that needs to be  taken into account is the fact that triplet-pairing SCs are likely to also exhibit a strong spin-orbit (SO) coupling. Thus it is worth discussing its effects on the topological phase diagrams. Intuitively, increasing the SO coupling is expected to decrease the localization length of Majorana bound states, and thus increase the size of the induced NW topological regions. 

In Fig.~\ref{phdSO} we show how the phase diagram presented in Fig.~\ref{phdlattice} changes when the substrate has a finite Rashba spin-orbit coupling. We introduce the latter into our model Hamiltonian in Eq.~(\ref{TBHam}) as follows
\begin{eqnarray*}
\mathcal{H}_{\mathrm{TB}} &= \sum\limits_{\bs{r}\in SC,\sigma} -\mu_{\mathrm{SC}} \Psi^\dag_{\bs{r}\sigma} \Psi_{\bs{r}\sigma} + \sigma \Delta_s \Psi_{\bs{r}\sigma} \Psi_{\bs{r},-\sigma} \\ 
 & - t\sum\limits_{\bs{r}\in SC,\sigma}\left( \Psi^\dag_{\bs{r}\sigma} \Psi_{\bs{r}+\bs{x},\sigma} +\Psi^\dag_{\bs{r}\sigma} \Psi_{\bs{r}+\bs{y},\sigma} \right)  \\
& + \lambda_{\mathrm{SC}} \negthickspace\sum\limits_{\bs{r}\in SC,\sigma} \left[  -\sigma \left(\Psi_{\bs{r}\sigma}^\dag \Psi_{\bs{r}+\bs{x},-\sigma} - \Psi_{\bs{r}\sigma} \Psi_{\bs{r}+\bs{x},-\sigma}^\dag \right) \right.\\
& \left. +i \left( \Psi_{\bs{r}\sigma}^\dag \Psi_{\bs{r}+\bs{y},-\sigma} - \Psi_{\bs{r}\sigma} \Psi_{\bs{r}+\bs{y},-\sigma}^\dag \right) \right]\\
& +\Delta_t \negthickspace \sum\limits_{\bs{r}\in SC,\sigma} \negthickspace \Psi_{\bs{r}\sigma} \Psi_{\bs{r}+\bs{x},\sigma} + i\sigma \Psi_{\bs{r}\sigma} \Psi_{\bs{r}+\bs{y},\sigma}\\
& +\sum\limits_{x\in NW,\sigma} -\mu_{\mathrm{NW}} c^\dag_{x\sigma} c_{x\sigma} - tc^\dag_{x\sigma} c_{x+1,\sigma}\\
& + t_{\mathrm{tun}}\sum\limits_{x \in NW,\sigma} \Psi^\dag_{x, y=0,\sigma} c_{x,\sigma} + \mathrm{H.c.},
\label{TBHamSO}
\end{eqnarray*}
where we introduced the spin-orbit coupling constant $\lambda_{\mathrm{SC}}$. In Fig.~\ref{phdSO}, $\lambda_{\mathrm{SC}}$ is set to $0$ on the upper panel, and to $0.2$ and $0.4$ in the middle and lower panels respectively. First, we note that the phase in which the NW is topological while the SC is trivial is achieved for a wider interval of parameters. In particular, Majorana Kramers pairs can form for lower values of the chemical potential $\mu_{\mathrm{NW}}$ of the NW. This can be intuitively understood by noting that larger spin-orbit couplings shorten the localization lengths for Majorana bound states and thus make them more stable. 

We note also that the critical $\Delta_s$ associated with the bulk phase transition (the black dashed lines in Fig.~\ref{phdSO}) is reduced in the presence of a finite SO coupling. To illustrate qualitatively this fact, we write the superconducting gap closing conditions established in Ref.~[\onlinecite{Sato2009}] in the following form:
\begin{align*}
\left[ - \mu_{\mathrm{SC}} - 2t \left(\cos p_{0x} + \cos p_{0y}  \right) \right]^2 &= \lambda_{\mathrm{SC}}^2 \left( \sin^2 p_{0x} + \sin^2 p_{0y}\right) \\
\Delta_t^2 \left( \sin^2 p_{0x} + \sin^2 p_{0y}\right) &= \Delta_s^2,
\end{align*}
where $(p_{0x}, p_{0y})$ is a set of points in the Brillouin zone where the SC gap closes. The first equation above defines the Fermi surface, whereas the second one is responsible for the gap closing. With a simple substitution we arrive at:
$$
\Delta_t^2 \left[ - \mu_{\mathrm{SC}} - 2t \left(\cos p_{0x} + \cos p_{0y}  \right) \right]^2 = \lambda_{\mathrm{SC}}^2 \Delta_s^2
$$
One can easily see that to achieve a gap closing point we need a smaller $\Delta_s$ in the presence of a finite SO coupling.

\section{The effects of disorder}

\begin{figure}
	\centering		
	\includegraphics[width=0.72\columnwidth]{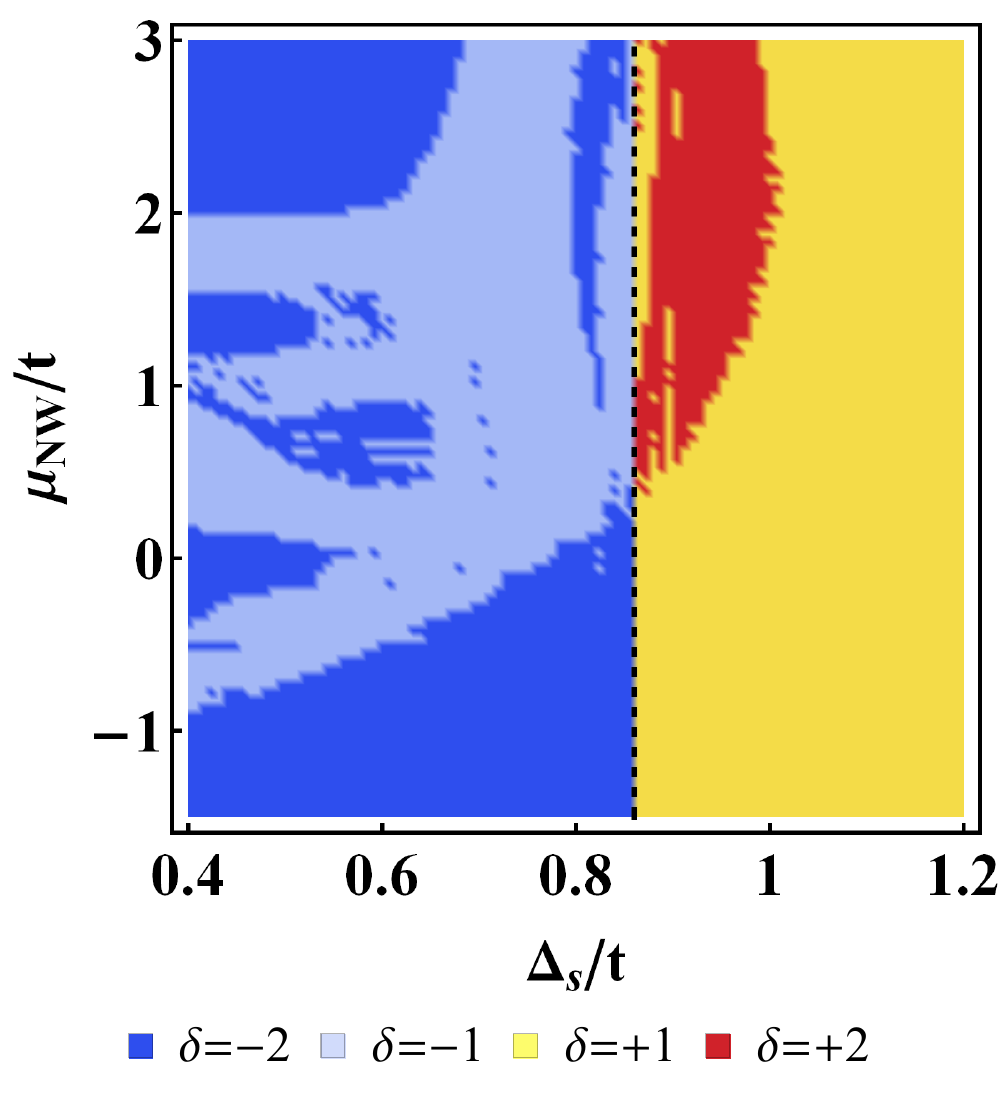}
	\caption{(Color online) The same phase diagram as in the upper panel of Fig.~\ref{phdSO} in the presence of disorder. The values of parameters are the same as in Fig.~\ref{phdlattice}.}
	\label{phdlatticedis}
\end{figure}

Another important issue to address is disorder. While reasonable amounts of potential disorder are not modifying considerably any key properties of conventional superconductors \cite{Anderson1959}, they may become important in unconventional superconductors not least due to formation of Shiba states and Shiba bands\cite{Balatsky2006}. The latter might lead to a significant gap renormalization in the substrate provided disorder potential is sufficiently strong, namely, greater than the hopping parameter $t$. However, in case of weak disorder impurity-induced states form very close to the superconducting gap edge\cite{Kaladzhyan2015}, and thus modify it only slightly while forming impurity-induced bands. Disorder could be considered along the same lines as e.g. in Refs.~[\onlinecite{Kaladzhyan2017a}],[\onlinecite{Kaladzhyan2017b}]. In this paper we model it as a random variation of the values of the chemical potentials $\mu_{NW}$ and $\mu_{SC}$ with an intensity of 2\% around their average values, both in the substrate and in the nanowire. We present the corresponding disordered phase diagram in Fig.~\ref{phdlatticedis}. As expected, the phase diagram is noisier than its disorder-free analog in the upper panel of Fig.~\ref{phdSO}, in other words, there are more regions with Majorana polarization smaller than $1$. Provided the system size is fixed, stronger disorder introduces too much noise into the phase diagram, eventually destroying the boundaries of topological phases. This happens due to the fact that it becomes easier for Majorana bound states to hybridize with each other in the presence of disorder.

\section{Discussion} 
In order to realize our proposal, we need a SC substrate with a mixed $s+p$-wave order parameter. 
Although there are some indirect evidences of a mixed order parameter in certain 2D TRS Pb-based superconductors 
\cite{Sekihara2013,Brun2014}, a 2D substrate with such property has not yet been fully established. Other options include SC doped topological materials \cite{Ando2015} or certain materials like $\mathrm{Sr_2RuO_4}$ which are believed to be triplet paired SCs \cite{Mackenzie2003,Mackenzie2017}. Recent experimental advances\cite{Wang2012,Xu2015} offer yet another possible route to realize mixed even and odd parity superconductivity via $s$-wave SC -- 3D topological insulator heterostructures.



As discussed above, in order to observe different topological phases in the NW and in the substrate the most important condition is to generate a Fermi momentum mismatch. This can be achieved  by tuning independently the chemical potentials in the wire and in the substrate. However, this can be experimentally challenging. Fortunately, a similar result can be obtained also by tuning a common chemical potential, if the band structure of the NW and the substrate are different, e.g. by choosing materials with very different effective masses, as shown in Fig.~\ref{figureA2} of Appendix \ref{AppendixAnalysis}.

In this work we have focused on the case of a TRS $p$-wave, also known as helical $p$-wave. However, our proposal  may apply also to a TRS-breaking $p$-wave SC, often referred to as "chiral $p$-wave". The latter gives rise to similar topological phase diagrams (see Appendix \ref{AppendixNonTRSpwave}). There is nevertheless, an important difference in this case -- the Majorana bound states pairs are not protected by TRS. However, we can establish that a
magnetic mirror symmetry \cite{Bernevig2014,Li2014,Andolina2017}, a weaker crystalline symmetry, can protect the Majorana Kramers pairs.\\

\section{Conclusion} We have demonstrated that a 1D nanowire proximitized by a 2D superconductor with mixed singlet and triplet pairing can enter a topological phase supporting Majorana Kramers pairs, even when the superconducting substrate is topologically trivial. 
More generally, we have shown that the topological phase diagram for the coupled system exhibits four different phases, such that both the SC substrate and the NW can be independently tuned in either a topological or a trivial phase.

Although we focus on a specific system, our findings are quite general and should be applicable also to various other configurations and dimensionalities. This is because the underlying physical mechanism relies on inducing an effective Fermi momentum in the wire different from that in the substrate. 

{\bf Acknowledgements}
We would like to acknowledge interesting discussions with Teemu Ojanen, Andrej Mesaros and Jens Bardarson. This work was partially supported  by the French Agence Nationale de la
Recherche through the contract ANR Mistral and the ERC Starting Independent Researcher Grant QUANTMATT $679722$.

\bibliography{biblio_integrating_out_s+p}

\begin{thebibliography}{42}%
\makeatletter
\providecommand \@ifxundefined [1]{%
 \@ifx{#1\undefined}
}%
\providecommand \@ifnum [1]{%
 \ifnum #1\expandafter \@firstoftwo
 \else \expandafter \@secondoftwo
 \fi
}%
\providecommand \@ifx [1]{%
 \ifx #1\expandafter \@firstoftwo
 \else \expandafter \@secondoftwo
 \fi
}%
\providecommand \natexlab [1]{#1}%
\providecommand \enquote  [1]{``#1''}%
\providecommand \bibnamefont  [1]{#1}%
\providecommand \bibfnamefont [1]{#1}%
\providecommand \citenamefont [1]{#1}%
\providecommand \href@noop [0]{\@secondoftwo}%
\providecommand \href [0]{\begingroup \@sanitize@url \@href}%
\providecommand \@href[1]{\@@startlink{#1}\@@href}%
\providecommand \@@href[1]{\endgroup#1\@@endlink}%
\providecommand \@sanitize@url [0]{\catcode `\\12\catcode `\$12\catcode
  `\&12\catcode `\#12\catcode `\^12\catcode `\_12\catcode `\%12\relax}%
\providecommand \@@startlink[1]{}%
\providecommand \@@endlink[0]{}%
\providecommand \url  [0]{\begingroup\@sanitize@url \@url }%
\providecommand \@url [1]{\endgroup\@href {#1}{\urlprefix }}%
\providecommand \urlprefix  [0]{URL }%
\providecommand \Eprint [0]{\href }%
\providecommand \doibase [0]{http://dx.doi.org/}%
\providecommand \selectlanguage [0]{\@gobble}%
\providecommand \bibinfo  [0]{\@secondoftwo}%
\providecommand \bibfield  [0]{\@secondoftwo}%
\providecommand \translation [1]{[#1]}%
\providecommand \BibitemOpen [0]{}%
\providecommand \bibitemStop [0]{}%
\providecommand \bibitemNoStop [0]{.\EOS\space}%
\providecommand \EOS [0]{\spacefactor3000\relax}%
\providecommand \BibitemShut  [1]{\csname bibitem#1\endcsname}%
\let\auto@bib@innerbib\@empty
\bibitem [{\citenamefont {Qi}\ \emph {et~al.}(2009)\citenamefont {Qi},
  \citenamefont {Hughes}, \citenamefont {Raghu},\ and\ \citenamefont
  {Zhang}}]{Qi2009}%
  \BibitemOpen
  \bibfield  {author} {\bibinfo {author} {\bibfnamefont {X.-L.}\ \bibnamefont
  {Qi}}, \bibinfo {author} {\bibfnamefont {T.~L.}\ \bibnamefont {Hughes}},
  \bibinfo {author} {\bibfnamefont {S.}~\bibnamefont {Raghu}}, \ and\ \bibinfo
  {author} {\bibfnamefont {S.-C.}\ \bibnamefont {Zhang}},\ }\href {\doibase
  10.1103/PhysRevLett.102.187001} {\bibfield  {journal} {\bibinfo  {journal}
  {Phys. Rev. Lett.}\ }\textbf {\bibinfo {volume} {102}},\ \bibinfo {pages}
  {187001} (\bibinfo {year} {2009})}\BibitemShut {NoStop}%
\bibitem [{\citenamefont {Qi}\ \emph {et~al.}(2010)\citenamefont {Qi},
  \citenamefont {Hughes},\ and\ \citenamefont {Zhang}}]{Qi2010}%
  \BibitemOpen
  \bibfield  {author} {\bibinfo {author} {\bibfnamefont {X.-L.}\ \bibnamefont
  {Qi}}, \bibinfo {author} {\bibfnamefont {T.~L.}\ \bibnamefont {Hughes}}, \
  and\ \bibinfo {author} {\bibfnamefont {S.-C.}\ \bibnamefont {Zhang}},\ }\href
  {\doibase 10.1103/PhysRevB.81.134508} {\bibfield  {journal} {\bibinfo
  {journal} {Phys. Rev. B}\ }\textbf {\bibinfo {volume} {81}},\ \bibinfo
  {pages} {134508} (\bibinfo {year} {2010})}\BibitemShut {NoStop}%
\bibitem [{\citenamefont {Deng}\ \emph {et~al.}(2012)\citenamefont {Deng},
  \citenamefont {Viola},\ and\ \citenamefont {Ortiz}}]{Deng2012}%
  \BibitemOpen
  \bibfield  {author} {\bibinfo {author} {\bibfnamefont {S.}~\bibnamefont
  {Deng}}, \bibinfo {author} {\bibfnamefont {L.}~\bibnamefont {Viola}}, \ and\
  \bibinfo {author} {\bibfnamefont {G.}~\bibnamefont {Ortiz}},\ }\href
  {\doibase 10.1103/PhysRevLett.108.036803} {\bibfield  {journal} {\bibinfo
  {journal} {Phys. Rev. Lett.}\ }\textbf {\bibinfo {volume} {108}},\ \bibinfo
  {pages} {036803} (\bibinfo {year} {2012})}\BibitemShut {NoStop}%
\bibitem [{\citenamefont {Wong}\ and\ \citenamefont {Law}(2012)}]{Wong2012}%
  \BibitemOpen
  \bibfield  {author} {\bibinfo {author} {\bibfnamefont {C.~L.~M.}\
  \bibnamefont {Wong}}\ and\ \bibinfo {author} {\bibfnamefont {K.~T.}\
  \bibnamefont {Law}},\ }\href {\doibase 10.1103/PhysRevB.86.184516} {\bibfield
   {journal} {\bibinfo  {journal} {Phys. Rev. B}\ }\textbf {\bibinfo {volume}
  {86}},\ \bibinfo {pages} {184516} (\bibinfo {year} {2012})}\BibitemShut
  {NoStop}%
\bibitem [{\citenamefont {Zhang}\ \emph {et~al.}(2013)\citenamefont {Zhang},
  \citenamefont {Kane},\ and\ \citenamefont {Mele}}]{Zhang2013}%
  \BibitemOpen
  \bibfield  {author} {\bibinfo {author} {\bibfnamefont {F.}~\bibnamefont
  {Zhang}}, \bibinfo {author} {\bibfnamefont {C.~L.}\ \bibnamefont {Kane}}, \
  and\ \bibinfo {author} {\bibfnamefont {E.~J.}\ \bibnamefont {Mele}},\ }\href
  {\doibase 10.1103/PhysRevLett.111.056402} {\bibfield  {journal} {\bibinfo
  {journal} {Phys. Rev. Lett.}\ }\textbf {\bibinfo {volume} {111}},\ \bibinfo
  {pages} {056402} (\bibinfo {year} {2013})}\BibitemShut {NoStop}%
\bibitem [{\citenamefont {Keselman}\ \emph {et~al.}(2013)\citenamefont
  {Keselman}, \citenamefont {Fu}, \citenamefont {Stern},\ and\ \citenamefont
  {Berg}}]{Keselman2013}%
  \BibitemOpen
  \bibfield  {author} {\bibinfo {author} {\bibfnamefont {A.}~\bibnamefont
  {Keselman}}, \bibinfo {author} {\bibfnamefont {L.}~\bibnamefont {Fu}},
  \bibinfo {author} {\bibfnamefont {A.}~\bibnamefont {Stern}}, \ and\ \bibinfo
  {author} {\bibfnamefont {E.}~\bibnamefont {Berg}},\ }\href {\doibase
  10.1103/PhysRevLett.111.116402} {\bibfield  {journal} {\bibinfo  {journal}
  {Phys. Rev. Lett.}\ }\textbf {\bibinfo {volume} {111}},\ \bibinfo {pages}
  {116402} (\bibinfo {year} {2013})}\BibitemShut {NoStop}%
\bibitem [{\citenamefont {Chung}\ \emph {et~al.}(2013)\citenamefont {Chung},
  \citenamefont {Horowitz},\ and\ \citenamefont {Qi}}]{Chung2013}%
  \BibitemOpen
  \bibfield  {author} {\bibinfo {author} {\bibfnamefont {S.~B.}\ \bibnamefont
  {Chung}}, \bibinfo {author} {\bibfnamefont {J.}~\bibnamefont {Horowitz}}, \
  and\ \bibinfo {author} {\bibfnamefont {X.-L.}\ \bibnamefont {Qi}},\ }\href
  {\doibase 10.1103/PhysRevB.88.214514} {\bibfield  {journal} {\bibinfo
  {journal} {Phys. Rev. B}\ }\textbf {\bibinfo {volume} {88}},\ \bibinfo
  {pages} {214514} (\bibinfo {year} {2013})}\BibitemShut {NoStop}%
\bibitem [{\citenamefont {Liu}\ \emph {et~al.}(2014)\citenamefont {Liu},
  \citenamefont {Wong},\ and\ \citenamefont {Law}}]{Liu2014}%
  \BibitemOpen
  \bibfield  {author} {\bibinfo {author} {\bibfnamefont {X.-J.}\ \bibnamefont
  {Liu}}, \bibinfo {author} {\bibfnamefont {C.~L.~M.}\ \bibnamefont {Wong}}, \
  and\ \bibinfo {author} {\bibfnamefont {K.~T.}\ \bibnamefont {Law}},\ }\href
  {\doibase 10.1103/PhysRevX.4.021018} {\bibfield  {journal} {\bibinfo
  {journal} {Phys. Rev. X}\ }\textbf {\bibinfo {volume} {4}},\ \bibinfo {pages}
  {021018} (\bibinfo {year} {2014})}\BibitemShut {NoStop}%
\bibitem [{\citenamefont {Gaidamauskas}\ \emph {et~al.}(2014)\citenamefont
  {Gaidamauskas}, \citenamefont {Paaske},\ and\ \citenamefont
  {Flensberg}}]{Gaidamauskas2014}%
  \BibitemOpen
  \bibfield  {author} {\bibinfo {author} {\bibfnamefont {E.}~\bibnamefont
  {Gaidamauskas}}, \bibinfo {author} {\bibfnamefont {J.}~\bibnamefont
  {Paaske}}, \ and\ \bibinfo {author} {\bibfnamefont {K.}~\bibnamefont
  {Flensberg}},\ }\href {\doibase 10.1103/PhysRevLett.112.126402} {\bibfield
  {journal} {\bibinfo  {journal} {Phys. Rev. Lett.}\ }\textbf {\bibinfo
  {volume} {112}},\ \bibinfo {pages} {126402} (\bibinfo {year}
  {2014})}\BibitemShut {NoStop}%
\bibitem [{\citenamefont {Dumitrescu}\ \emph {et~al.}(2014)\citenamefont
  {Dumitrescu}, \citenamefont {Sau},\ and\ \citenamefont
  {Tewari}}]{Dumitrescu2014}%
  \BibitemOpen
  \bibfield  {author} {\bibinfo {author} {\bibfnamefont {E.}~\bibnamefont
  {Dumitrescu}}, \bibinfo {author} {\bibfnamefont {J.~D.}\ \bibnamefont {Sau}},
  \ and\ \bibinfo {author} {\bibfnamefont {S.}~\bibnamefont {Tewari}},\ }\href
  {\doibase 10.1103/PhysRevB.90.245438} {\bibfield  {journal} {\bibinfo
  {journal} {Phys. Rev. B}\ }\textbf {\bibinfo {volume} {90}},\ \bibinfo
  {pages} {245438} (\bibinfo {year} {2014})}\BibitemShut {NoStop}%
\bibitem [{\citenamefont {Haim}\ \emph {et~al.}(2014)\citenamefont {Haim},
  \citenamefont {Keselman}, \citenamefont {Berg},\ and\ \citenamefont
  {Oreg}}]{Haim2014}%
  \BibitemOpen
  \bibfield  {author} {\bibinfo {author} {\bibfnamefont {A.}~\bibnamefont
  {Haim}}, \bibinfo {author} {\bibfnamefont {A.}~\bibnamefont {Keselman}},
  \bibinfo {author} {\bibfnamefont {E.}~\bibnamefont {Berg}}, \ and\ \bibinfo
  {author} {\bibfnamefont {Y.}~\bibnamefont {Oreg}},\ }\href {\doibase
  10.1103/PhysRevB.89.220504} {\bibfield  {journal} {\bibinfo  {journal} {Phys.
  Rev. B}\ }\textbf {\bibinfo {volume} {89}},\ \bibinfo {pages} {220504}
  (\bibinfo {year} {2014})}\BibitemShut {NoStop}%
\bibitem [{\citenamefont {Klinovaja}\ \emph {et~al.}(2014)\citenamefont
  {Klinovaja}, \citenamefont {Yacoby},\ and\ \citenamefont
  {Loss}}]{Klinovaja2014}%
  \BibitemOpen
  \bibfield  {author} {\bibinfo {author} {\bibfnamefont {J.}~\bibnamefont
  {Klinovaja}}, \bibinfo {author} {\bibfnamefont {A.}~\bibnamefont {Yacoby}}, \
  and\ \bibinfo {author} {\bibfnamefont {D.}~\bibnamefont {Loss}},\ }\href
  {\doibase 10.1103/PhysRevB.90.155447} {\bibfield  {journal} {\bibinfo
  {journal} {Phys. Rev. B}\ }\textbf {\bibinfo {volume} {90}},\ \bibinfo
  {pages} {155447} (\bibinfo {year} {2014})}\BibitemShut {NoStop}%
\bibitem [{\citenamefont {Klinovaja}\ and\ \citenamefont
  {Loss}(2014)}]{Klinovaja2014-2}%
  \BibitemOpen
  \bibfield  {author} {\bibinfo {author} {\bibfnamefont {J.}~\bibnamefont
  {Klinovaja}}\ and\ \bibinfo {author} {\bibfnamefont {D.}~\bibnamefont
  {Loss}},\ }\href {\doibase 10.1103/PhysRevB.90.045118} {\bibfield  {journal}
  {\bibinfo  {journal} {Phys. Rev. B}\ }\textbf {\bibinfo {volume} {90}},\
  \bibinfo {pages} {045118} (\bibinfo {year} {2014})}\BibitemShut {NoStop}%
\bibitem [{\citenamefont {Chevallier}\ \emph {et~al.}(2013)\citenamefont
  {Chevallier}, \citenamefont {Simon},\ and\ \citenamefont
  {Bena}}]{Chevallier2013}%
  \BibitemOpen
  \bibfield  {author} {\bibinfo {author} {\bibfnamefont {D.}~\bibnamefont
  {Chevallier}}, \bibinfo {author} {\bibfnamefont {P.}~\bibnamefont {Simon}}, \
  and\ \bibinfo {author} {\bibfnamefont {C.}~\bibnamefont {Bena}},\ }\href
  {\doibase 10.1103/PhysRevB.88.165401} {\bibfield  {journal} {\bibinfo
  {journal} {Phys. Rev. B}\ }\textbf {\bibinfo {volume} {88}},\ \bibinfo
  {pages} {165401} (\bibinfo {year} {2013})}\BibitemShut {NoStop}%
\bibitem [{\citenamefont {Nakosai}\ \emph {et~al.}(2013)\citenamefont
  {Nakosai}, \citenamefont {Budich}, \citenamefont {Tanaka}, \citenamefont
  {Trauzettel},\ and\ \citenamefont {Nagaosa}}]{Nakosai2013prl}%
  \BibitemOpen
  \bibfield  {author} {\bibinfo {author} {\bibfnamefont {S.}~\bibnamefont
  {Nakosai}}, \bibinfo {author} {\bibfnamefont {J.~C.}\ \bibnamefont {Budich}},
  \bibinfo {author} {\bibfnamefont {Y.}~\bibnamefont {Tanaka}}, \bibinfo
  {author} {\bibfnamefont {B.}~\bibnamefont {Trauzettel}}, \ and\ \bibinfo
  {author} {\bibfnamefont {N.}~\bibnamefont {Nagaosa}},\ }\href {\doibase
  10.1103/PhysRevLett.110.117002} {\bibfield  {journal} {\bibinfo  {journal}
  {Phys. Rev. Lett.}\ }\textbf {\bibinfo {volume} {110}},\ \bibinfo {pages}
  {117002} (\bibinfo {year} {2013})}\BibitemShut {NoStop}%
\bibitem [{\citenamefont {Neupert}\ \emph {et~al.}(2016)\citenamefont
  {Neupert}, \citenamefont {Yazdani},\ and\ \citenamefont
  {Bernevig}}]{Neupert2016}%
  \BibitemOpen
  \bibfield  {author} {\bibinfo {author} {\bibfnamefont {T.}~\bibnamefont
  {Neupert}}, \bibinfo {author} {\bibfnamefont {A.}~\bibnamefont {Yazdani}}, \
  and\ \bibinfo {author} {\bibfnamefont {B.~A.}\ \bibnamefont {Bernevig}},\
  }\href {\doibase 10.1103/PhysRevB.93.094508} {\bibfield  {journal} {\bibinfo
  {journal} {Phys. Rev. B}\ }\textbf {\bibinfo {volume} {93}},\ \bibinfo
  {pages} {094508} (\bibinfo {year} {2016})}\BibitemShut {NoStop}%
\bibitem [{\citenamefont {Yu}(1965)}]{Yu1965}%
  \BibitemOpen
  \bibfield  {author} {\bibinfo {author} {\bibfnamefont {L.}~\bibnamefont
  {Yu}},\ }\href {\doibase 10.7498/aps.21.75} {\bibfield  {journal} {\bibinfo
  {journal} {Acta Physica Sinica}\ }\textbf {\bibinfo {volume} {21}},\ \bibinfo
  {pages} {75} (\bibinfo {year} {1965})}\BibitemShut {NoStop}%
\bibitem [{\citenamefont {Shiba}(1968)}]{Shiba1968}%
  \BibitemOpen
  \bibfield  {author} {\bibinfo {author} {\bibfnamefont {H.}~\bibnamefont
  {Shiba}},\ }\href {\doibase 10.1143/PTP.40.435} {\bibfield  {journal}
  {\bibinfo  {journal} {Progress of Theoretical Physics}\ }\textbf {\bibinfo
  {volume} {40}},\ \bibinfo {pages} {435} (\bibinfo {year} {1968})}\BibitemShut
  {NoStop}%
\bibitem [{\citenamefont {Rusinov}(1969)}]{Rusinov1969}%
  \BibitemOpen
  \bibfield  {author} {\bibinfo {author} {\bibfnamefont {A.~I.}\ \bibnamefont
  {Rusinov}},\ }\href@noop {} {\bibfield  {journal} {\bibinfo  {journal}
  {Soviet Journal of Experimental and Theoretical Physics Letters}\ }\textbf
  {\bibinfo {volume} {9}} (\bibinfo {year} {1969})}\BibitemShut {NoStop}%
\bibitem [{\citenamefont {Anderson}(1959)}]{Anderson1959}%
  \BibitemOpen
  \bibfield  {author} {\bibinfo {author} {\bibfnamefont {P.}~\bibnamefont
  {Anderson}},\ }\href {\doibase https://doi.org/10.1016/0022-3697(59)90036-8}
  {\bibfield  {journal} {\bibinfo  {journal} {Journal of Physics and Chemistry
  of Solids}\ }\textbf {\bibinfo {volume} {11}},\ \bibinfo {pages} {26}
  (\bibinfo {year} {1959})}\BibitemShut {NoStop}%
\bibitem [{\citenamefont {Balatsky}\ \emph {et~al.}(2006)\citenamefont
  {Balatsky}, \citenamefont {Vekhter},\ and\ \citenamefont
  {Zhu}}]{Balatsky2006}%
  \BibitemOpen
  \bibfield  {author} {\bibinfo {author} {\bibfnamefont {A.~V.}\ \bibnamefont
  {Balatsky}}, \bibinfo {author} {\bibfnamefont {I.}~\bibnamefont {Vekhter}}, \
  and\ \bibinfo {author} {\bibfnamefont {J.-X.}\ \bibnamefont {Zhu}},\ }\href
  {\doibase 10.1103/RevModPhys.78.373} {\bibfield  {journal} {\bibinfo
  {journal} {Rev. Mod. Phys.}\ }\textbf {\bibinfo {volume} {78}},\ \bibinfo
  {pages} {373} (\bibinfo {year} {2006})}\BibitemShut {NoStop}%
\bibitem [{\citenamefont {Kaladzhyan}\ \emph
  {et~al.}(2016{\natexlab{a}})\citenamefont {Kaladzhyan}, \citenamefont
  {Bena},\ and\ \citenamefont {Simon}}]{Kaladzhyan2015}%
  \BibitemOpen
  \bibfield  {author} {\bibinfo {author} {\bibfnamefont {V.}~\bibnamefont
  {Kaladzhyan}}, \bibinfo {author} {\bibfnamefont {C.}~\bibnamefont {Bena}}, \
  and\ \bibinfo {author} {\bibfnamefont {P.}~\bibnamefont {Simon}},\ }\href
  {\doibase 10.1103/PhysRevB.93.214514} {\bibfield  {journal} {\bibinfo
  {journal} {Phys. Rev. B}\ }\textbf {\bibinfo {volume} {93}},\ \bibinfo
  {pages} {214514} (\bibinfo {year} {2016}{\natexlab{a}})}\BibitemShut
  {NoStop}%
\bibitem [{\citenamefont {Kaladzhyan}\ \emph
  {et~al.}(2016{\natexlab{b}})\citenamefont {Kaladzhyan}, \citenamefont
  {R\"ontynen}, \citenamefont {Simon},\ and\ \citenamefont
  {Ojanen}}]{Kaladzhyan2016a}%
  \BibitemOpen
  \bibfield  {author} {\bibinfo {author} {\bibfnamefont {V.}~\bibnamefont
  {Kaladzhyan}}, \bibinfo {author} {\bibfnamefont {J.}~\bibnamefont
  {R\"ontynen}}, \bibinfo {author} {\bibfnamefont {P.}~\bibnamefont {Simon}}, \
  and\ \bibinfo {author} {\bibfnamefont {T.}~\bibnamefont {Ojanen}},\ }\href
  {\doibase 10.1103/PhysRevB.94.060505} {\bibfield  {journal} {\bibinfo
  {journal} {Phys. Rev. B}\ }\textbf {\bibinfo {volume} {94}},\ \bibinfo
  {pages} {060505} (\bibinfo {year} {2016}{\natexlab{b}})}\BibitemShut
  {NoStop}%
\bibitem [{\citenamefont {Hsieh}\ \emph {et~al.}(2016)\citenamefont {Hsieh},
  \citenamefont {Ishizuka}, \citenamefont {Balents},\ and\ \citenamefont
  {Hughes}}]{Hsieh2016}%
  \BibitemOpen
  \bibfield  {author} {\bibinfo {author} {\bibfnamefont {T.~H.}\ \bibnamefont
  {Hsieh}}, \bibinfo {author} {\bibfnamefont {H.}~\bibnamefont {Ishizuka}},
  \bibinfo {author} {\bibfnamefont {L.}~\bibnamefont {Balents}}, \ and\
  \bibinfo {author} {\bibfnamefont {T.~L.}\ \bibnamefont {Hughes}},\ }\href
  {\doibase 10.1103/PhysRevLett.116.086802} {\bibfield  {journal} {\bibinfo
  {journal} {Phys. Rev. Lett.}\ }\textbf {\bibinfo {volume} {116}},\ \bibinfo
  {pages} {086802} (\bibinfo {year} {2016})}\BibitemShut {NoStop}%
\bibitem [{Note1()}]{Note1}%
  \BibitemOpen
  \bibinfo {note} {Even though this value is close to the actual value of the
  induced gap, we should note that this expression is however an approximation
  valid only when the spectrum can be linearized.}\BibitemShut {Stop}%
\bibitem [{\citenamefont {Sato}\ and\ \citenamefont
  {Fujimoto}(2009)}]{Sato2009}%
  \BibitemOpen
  \bibfield  {author} {\bibinfo {author} {\bibfnamefont {M.}~\bibnamefont
  {Sato}}\ and\ \bibinfo {author} {\bibfnamefont {S.}~\bibnamefont
  {Fujimoto}},\ }\href {\doibase 10.1103/PhysRevB.79.094504} {\bibfield
  {journal} {\bibinfo  {journal} {Phys. Rev. B}\ }\textbf {\bibinfo {volume}
  {79}},\ \bibinfo {pages} {094504} (\bibinfo {year} {2009})}\BibitemShut
  {NoStop}%
\bibitem [{mat()}]{matq}%
  \BibitemOpen
  \href@noop {} {\emph {\bibinfo {title} {MatQ}}},\ \bibinfo {address}
  {www.icmm.csic.es/sanjose/MathQ/MathQ.html}\BibitemShut {NoStop}%
\bibitem [{\citenamefont {Sticlet}\ \emph {et~al.}(2012)\citenamefont
  {Sticlet}, \citenamefont {Bena},\ and\ \citenamefont {Simon}}]{Sticlet2012}%
  \BibitemOpen
  \bibfield  {author} {\bibinfo {author} {\bibfnamefont {D.}~\bibnamefont
  {Sticlet}}, \bibinfo {author} {\bibfnamefont {C.}~\bibnamefont {Bena}}, \
  and\ \bibinfo {author} {\bibfnamefont {P.}~\bibnamefont {Simon}},\ }\href
  {\doibase 10.1103/PhysRevLett.108.096802} {\bibfield  {journal} {\bibinfo
  {journal} {Phys. Rev. Lett.}\ }\textbf {\bibinfo {volume} {108}},\ \bibinfo
  {pages} {096802} (\bibinfo {year} {2012})}\BibitemShut {NoStop}%
\bibitem [{\citenamefont {Sedlmayr}\ and\ \citenamefont
  {Bena}(2015)}]{Sedlmayr2015b}%
  \BibitemOpen
  \bibfield  {author} {\bibinfo {author} {\bibfnamefont {N.}~\bibnamefont
  {Sedlmayr}}\ and\ \bibinfo {author} {\bibfnamefont {C.}~\bibnamefont
  {Bena}},\ }\href {\doibase 10.1103/PhysRevB.92.115115} {\bibfield  {journal}
  {\bibinfo  {journal} {Phys. Rev. B}\ }\textbf {\bibinfo {volume} {92}},\
  \bibinfo {pages} {115115} (\bibinfo {year} {2015})}\BibitemShut {NoStop}%
\bibitem [{\citenamefont {Sedlmayr}\ \emph {et~al.}(2016)\citenamefont
  {Sedlmayr}, \citenamefont {Aguiar-Hualde},\ and\ \citenamefont
  {Bena}}]{Sedlmayr2016}%
  \BibitemOpen
  \bibfield  {author} {\bibinfo {author} {\bibfnamefont {N.}~\bibnamefont
  {Sedlmayr}}, \bibinfo {author} {\bibfnamefont {J.~M.}\ \bibnamefont
  {Aguiar-Hualde}}, \ and\ \bibinfo {author} {\bibfnamefont {C.}~\bibnamefont
  {Bena}},\ }\href {\doibase 10.1103/PhysRevB.93.155425} {\bibfield  {journal}
  {\bibinfo  {journal} {Phys. Rev. B}\ }\textbf {\bibinfo {volume} {93}},\
  \bibinfo {pages} {155425} (\bibinfo {year} {2016})}\BibitemShut {NoStop}%
\bibitem [{\citenamefont {Kaladzhyan}\ and\ \citenamefont
  {Bena}(2017)}]{Kaladzhyan2017a}%
  \BibitemOpen
  \bibfield  {author} {\bibinfo {author} {\bibfnamefont {V.}~\bibnamefont
  {Kaladzhyan}}\ and\ \bibinfo {author} {\bibfnamefont {C.}~\bibnamefont
  {Bena}},\ }\href {\doibase 10.21468/SciPostPhys.3.1.002} {\bibfield
  {journal} {\bibinfo  {journal} {SciPost Phys.}\ }\textbf {\bibinfo {volume}
  {3}},\ \bibinfo {pages} {002} (\bibinfo {year} {2017})}\BibitemShut {NoStop}%
\bibitem [{\citenamefont {Kaladzhyan}\ \emph {et~al.}(2017)\citenamefont
  {Kaladzhyan}, \citenamefont {Despres}, \citenamefont {Mandal},\ and\
  \citenamefont {Bena}}]{Kaladzhyan2017b}%
  \BibitemOpen
  \bibfield  {author} {\bibinfo {author} {\bibfnamefont {V.}~\bibnamefont
  {Kaladzhyan}}, \bibinfo {author} {\bibfnamefont {J.}~\bibnamefont {Despres}},
  \bibinfo {author} {\bibfnamefont {I.}~\bibnamefont {Mandal}}, \ and\ \bibinfo
  {author} {\bibfnamefont {C.}~\bibnamefont {Bena}},\ }\href {\doibase
  10.1140/epjb/e2017-80103-y} {\bibfield  {journal} {\bibinfo  {journal} {The
  European Physical Journal B}\ }\textbf {\bibinfo {volume} {90}},\ \bibinfo
  {pages} {211} (\bibinfo {year} {2017})}\BibitemShut {NoStop}%
\bibitem [{\citenamefont {Sekihara}\ \emph {et~al.}(2013)\citenamefont
  {Sekihara}, \citenamefont {Masutomi},\ and\ \citenamefont
  {Okamoto}}]{Sekihara2013}%
  \BibitemOpen
  \bibfield  {author} {\bibinfo {author} {\bibfnamefont {T.}~\bibnamefont
  {Sekihara}}, \bibinfo {author} {\bibfnamefont {R.}~\bibnamefont {Masutomi}},
  \ and\ \bibinfo {author} {\bibfnamefont {T.}~\bibnamefont {Okamoto}},\ }\href
  {\doibase 10.1103/PhysRevLett.111.057005} {\bibfield  {journal} {\bibinfo
  {journal} {Phys. Rev. Lett.}\ }\textbf {\bibinfo {volume} {111}},\ \bibinfo
  {pages} {057005} (\bibinfo {year} {2013})}\BibitemShut {NoStop}%
\bibitem [{\citenamefont {Brun}\ and\ \citenamefont {et~al.}(2014)}]{Brun2014}%
  \BibitemOpen
  \bibfield  {author} {\bibinfo {author} {\bibfnamefont {C.}~\bibnamefont
  {Brun}}\ and\ \bibinfo {author} {\bibnamefont {et~al.}},\ }\href@noop {}
  {\bibfield  {journal} {\bibinfo  {journal} {Nature Physics}\ }\textbf
  {\bibinfo {volume} {10}},\ \bibinfo {pages} {444} (\bibinfo {year}
  {2014})}\BibitemShut {NoStop}%
\bibitem [{\citenamefont {Ando}\ and\ \citenamefont {Fu}(2015)}]{Ando2015}%
  \BibitemOpen
  \bibfield  {author} {\bibinfo {author} {\bibfnamefont {Y.}~\bibnamefont
  {Ando}}\ and\ \bibinfo {author} {\bibfnamefont {L.}~\bibnamefont {Fu}},\
  }\href {\doibase 10.1146/annurev-conmatphys-031214-014501} {\bibfield
  {journal} {\bibinfo  {journal} {Annual Rev. of Cond. Matt. Physics}\ }\textbf
  {\bibinfo {volume} {6}},\ \bibinfo {pages} {361} (\bibinfo {year}
  {2015})}\BibitemShut {NoStop}%
\bibitem [{\citenamefont {Mackenzie}\ and\ \citenamefont
  {Maeno}(2003)}]{Mackenzie2003}%
  \BibitemOpen
  \bibfield  {author} {\bibinfo {author} {\bibfnamefont {A.~P.}\ \bibnamefont
  {Mackenzie}}\ and\ \bibinfo {author} {\bibfnamefont {Y.}~\bibnamefont
  {Maeno}},\ }\href {\doibase 10.1103/RevModPhys.75.657} {\bibfield  {journal}
  {\bibinfo  {journal} {Rev. Mod. Phys.}\ }\textbf {\bibinfo {volume} {75}},\
  \bibinfo {pages} {657} (\bibinfo {year} {2003})}\BibitemShut {NoStop}%
\bibitem [{\citenamefont {Mackenzie}\ \emph {et~al.}(2017)\citenamefont
  {Mackenzie}, \citenamefont {Thomas~Scaffidi}, \citenamefont {Hicks},\ and\
  \citenamefont {Maeno}}]{Mackenzie2017}%
  \BibitemOpen
  \bibfield  {author} {\bibinfo {author} {\bibfnamefont {A.}~\bibnamefont
  {Mackenzie}}, \bibinfo {author} {\bibfnamefont {T.}~\bibnamefont
  {Thomas~Scaffidi}}, \bibinfo {author} {\bibfnamefont {C.}~\bibnamefont
  {Hicks}}, \ and\ \bibinfo {author} {\bibfnamefont {Y.}~\bibnamefont
  {Maeno}},\ }\href@noop {} {\bibfield  {journal} {\bibinfo  {journal}
  {arXiv:1706.01942}\ } (\bibinfo {year} {2017})}\BibitemShut {NoStop}%
\bibitem [{\citenamefont {Wang}\ \emph {et~al.}(2012)\citenamefont {Wang},
  \citenamefont {Liu}, \citenamefont {Xu}, \citenamefont {Yang}, \citenamefont
  {Miao}, \citenamefont {Yao}, \citenamefont {Gao}, \citenamefont {Shen},
  \citenamefont {Ma}, \citenamefont {Chen}, \citenamefont {Xu}, \citenamefont
  {Liu}, \citenamefont {Zhang}, \citenamefont {Qian}, \citenamefont {Jia},\
  and\ \citenamefont {Xue}}]{Wang2012}%
  \BibitemOpen
  \bibfield  {author} {\bibinfo {author} {\bibfnamefont {M.-X.}\ \bibnamefont
  {Wang}}, \bibinfo {author} {\bibfnamefont {C.}~\bibnamefont {Liu}}, \bibinfo
  {author} {\bibfnamefont {J.-P.}\ \bibnamefont {Xu}}, \bibinfo {author}
  {\bibfnamefont {F.}~\bibnamefont {Yang}}, \bibinfo {author} {\bibfnamefont
  {L.}~\bibnamefont {Miao}}, \bibinfo {author} {\bibfnamefont {M.-Y.}\
  \bibnamefont {Yao}}, \bibinfo {author} {\bibfnamefont {C.~L.}\ \bibnamefont
  {Gao}}, \bibinfo {author} {\bibfnamefont {C.}~\bibnamefont {Shen}}, \bibinfo
  {author} {\bibfnamefont {X.}~\bibnamefont {Ma}}, \bibinfo {author}
  {\bibfnamefont {X.}~\bibnamefont {Chen}}, \bibinfo {author} {\bibfnamefont
  {Z.-A.}\ \bibnamefont {Xu}}, \bibinfo {author} {\bibfnamefont
  {Y.}~\bibnamefont {Liu}}, \bibinfo {author} {\bibfnamefont {S.-C.}\
  \bibnamefont {Zhang}}, \bibinfo {author} {\bibfnamefont {D.}~\bibnamefont
  {Qian}}, \bibinfo {author} {\bibfnamefont {J.-F.}\ \bibnamefont {Jia}}, \
  and\ \bibinfo {author} {\bibfnamefont {Q.-K.}\ \bibnamefont {Xue}},\ }\href
  {\doibase 10.1126/science.1216466} {\bibfield  {journal} {\bibinfo  {journal}
  {Science}\ }\textbf {\bibinfo {volume} {336}},\ \bibinfo {pages} {52}
  (\bibinfo {year} {2012})}\BibitemShut {NoStop}%
\bibitem [{\citenamefont {Xu}\ \emph {et~al.}(2015)\citenamefont {Xu},
  \citenamefont {Wang}, \citenamefont {Liu}, \citenamefont {Ge}, \citenamefont
  {Yang}, \citenamefont {Liu}, \citenamefont {Xu}, \citenamefont {Guan},
  \citenamefont {Gao}, \citenamefont {Qian}, \citenamefont {Liu}, \citenamefont
  {Wang}, \citenamefont {Zhang}, \citenamefont {Xue},\ and\ \citenamefont
  {Jia}}]{Xu2015}%
  \BibitemOpen
  \bibfield  {author} {\bibinfo {author} {\bibfnamefont {J.-P.}\ \bibnamefont
  {Xu}}, \bibinfo {author} {\bibfnamefont {M.-X.}\ \bibnamefont {Wang}},
  \bibinfo {author} {\bibfnamefont {Z.~L.}\ \bibnamefont {Liu}}, \bibinfo
  {author} {\bibfnamefont {J.-F.}\ \bibnamefont {Ge}}, \bibinfo {author}
  {\bibfnamefont {X.}~\bibnamefont {Yang}}, \bibinfo {author} {\bibfnamefont
  {C.}~\bibnamefont {Liu}}, \bibinfo {author} {\bibfnamefont {Z.~A.}\
  \bibnamefont {Xu}}, \bibinfo {author} {\bibfnamefont {D.}~\bibnamefont
  {Guan}}, \bibinfo {author} {\bibfnamefont {C.~L.}\ \bibnamefont {Gao}},
  \bibinfo {author} {\bibfnamefont {D.}~\bibnamefont {Qian}}, \bibinfo {author}
  {\bibfnamefont {Y.}~\bibnamefont {Liu}}, \bibinfo {author} {\bibfnamefont
  {Q.-H.}\ \bibnamefont {Wang}}, \bibinfo {author} {\bibfnamefont {F.-C.}\
  \bibnamefont {Zhang}}, \bibinfo {author} {\bibfnamefont {Q.-K.}\ \bibnamefont
  {Xue}}, \ and\ \bibinfo {author} {\bibfnamefont {J.-F.}\ \bibnamefont
  {Jia}},\ }\href {\doibase 10.1103/PhysRevLett.114.017001} {\bibfield
  {journal} {\bibinfo  {journal} {Phys. Rev. Lett.}\ }\textbf {\bibinfo
  {volume} {114}},\ \bibinfo {pages} {017001} (\bibinfo {year}
  {2015})}\BibitemShut {NoStop}%
\bibitem [{\citenamefont {Fang}\ \emph {et~al.}(2014)\citenamefont {Fang},
  \citenamefont {Gilbert},\ and\ \citenamefont {Bernevig}}]{Bernevig2014}%
  \BibitemOpen
  \bibfield  {author} {\bibinfo {author} {\bibfnamefont {C.}~\bibnamefont
  {Fang}}, \bibinfo {author} {\bibfnamefont {M.~J.}\ \bibnamefont {Gilbert}}, \
  and\ \bibinfo {author} {\bibfnamefont {B.~A.}\ \bibnamefont {Bernevig}},\
  }\href {\doibase 10.1103/PhysRevLett.112.106401} {\bibfield  {journal}
  {\bibinfo  {journal} {Phys. Rev. Lett.}\ }\textbf {\bibinfo {volume} {112}},\
  \bibinfo {pages} {106401} (\bibinfo {year} {2014})}\BibitemShut {NoStop}%
\bibitem [{\citenamefont {Li}\ \emph {et~al.}(2014)\citenamefont {Li},
  \citenamefont {Chen}, \citenamefont {Drozdov}, \citenamefont {Yazdani},
  \citenamefont {Bernevig},\ and\ \citenamefont {MacDonald}}]{Li2014}%
  \BibitemOpen
  \bibfield  {author} {\bibinfo {author} {\bibfnamefont {J.}~\bibnamefont
  {Li}}, \bibinfo {author} {\bibfnamefont {H.}~\bibnamefont {Chen}}, \bibinfo
  {author} {\bibfnamefont {I.~K.}\ \bibnamefont {Drozdov}}, \bibinfo {author}
  {\bibfnamefont {A.}~\bibnamefont {Yazdani}}, \bibinfo {author} {\bibfnamefont
  {B.~A.}\ \bibnamefont {Bernevig}}, \ and\ \bibinfo {author} {\bibfnamefont
  {A.~H.}\ \bibnamefont {MacDonald}},\ }\href {\doibase
  10.1103/PhysRevB.90.235433} {\bibfield  {journal} {\bibinfo  {journal} {Phys.
  Rev. B}\ }\textbf {\bibinfo {volume} {90}},\ \bibinfo {pages} {235433}
  (\bibinfo {year} {2014})}\BibitemShut {NoStop}%
\bibitem [{\citenamefont {Andolina}\ and\ \citenamefont
  {Simon}(2017)}]{Andolina2017}%
  \BibitemOpen
  \bibfield  {author} {\bibinfo {author} {\bibfnamefont {G.~M.}\ \bibnamefont
  {Andolina}}\ and\ \bibinfo {author} {\bibfnamefont {P.}~\bibnamefont
  {Simon}},\ }\href {\doibase 10.1103/PhysRevB.96.235411} {\bibfield  {journal}
  {\bibinfo  {journal} {Phys. Rev. B}\ }\textbf {\bibinfo {volume} {96}},\
  \bibinfo {pages} {235411} (\bibinfo {year} {2017})}\BibitemShut {NoStop}%
\end{thebibliography}%

\newpage
\widetext
\appendix   
\renewcommand{\thefigure}{A\arabic{figure}}
\setcounter{section}{0}
\setcounter{figure}{0}
\setcounter{equation}{0}

\section{Comments on inducing $S+P$--wave superconductivity via proximity effect}\label{AppendixAnalysis}

\begin{figure}
	\centering
	\includegraphics[width=0.47\columnwidth]{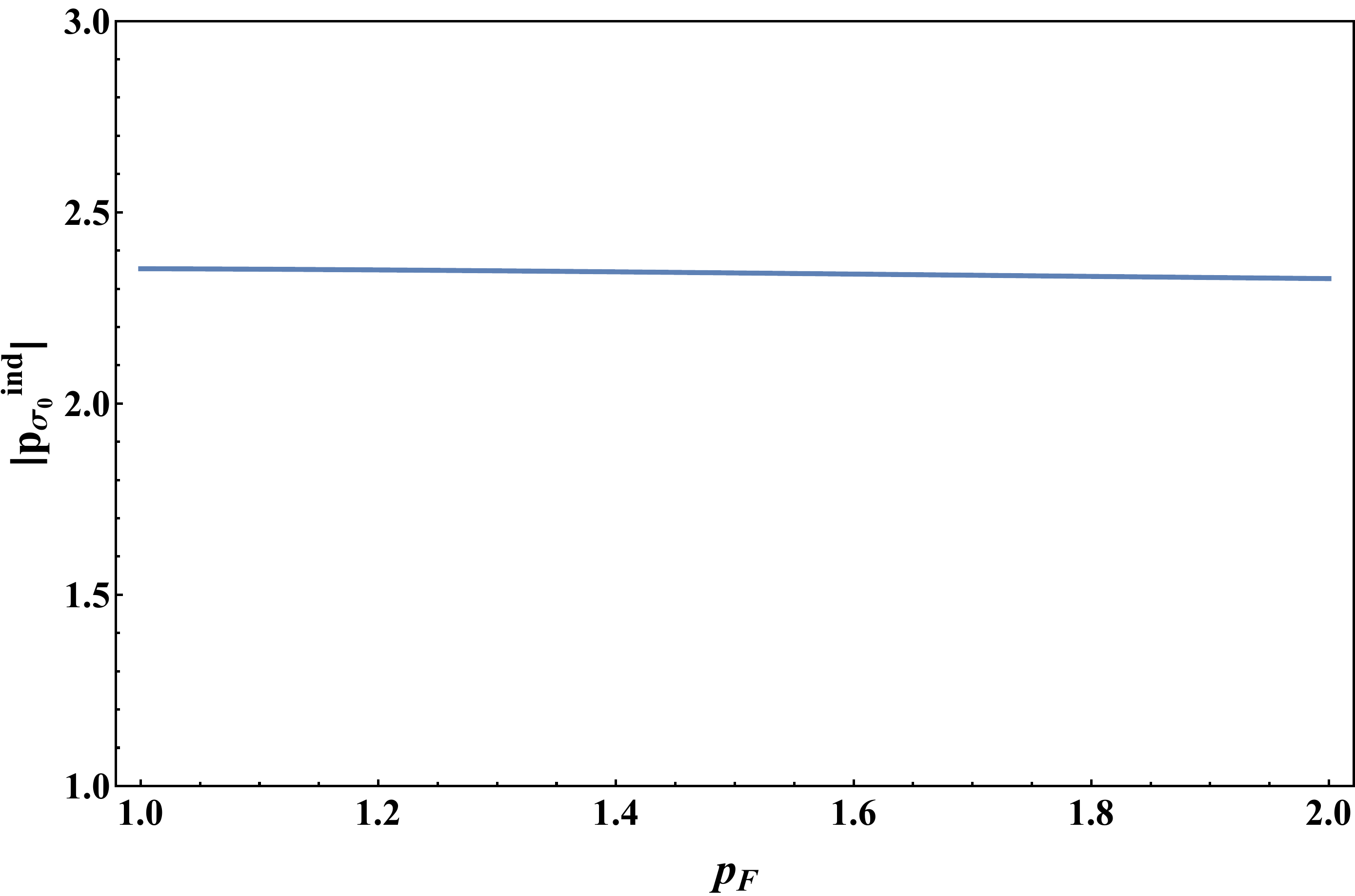}	
	\caption{(Color online) The Fermi momentum of the NW plotted as a function of the Fermi momentum of the substrate. As we see from this plot, it is almost independent on the Fermi momentum of the substrate. We took $\Delta_s = \varkappa(0.1+p_F)$, and we have set $M=m=1.5, \varkappa=2.5, \mu=2$, and $t_{\mathrm{tun}}=0.5$.}
	\label{figureA1}
\end{figure}

\begin{figure}
	\centering		
	\includegraphics[width=0.47\columnwidth]{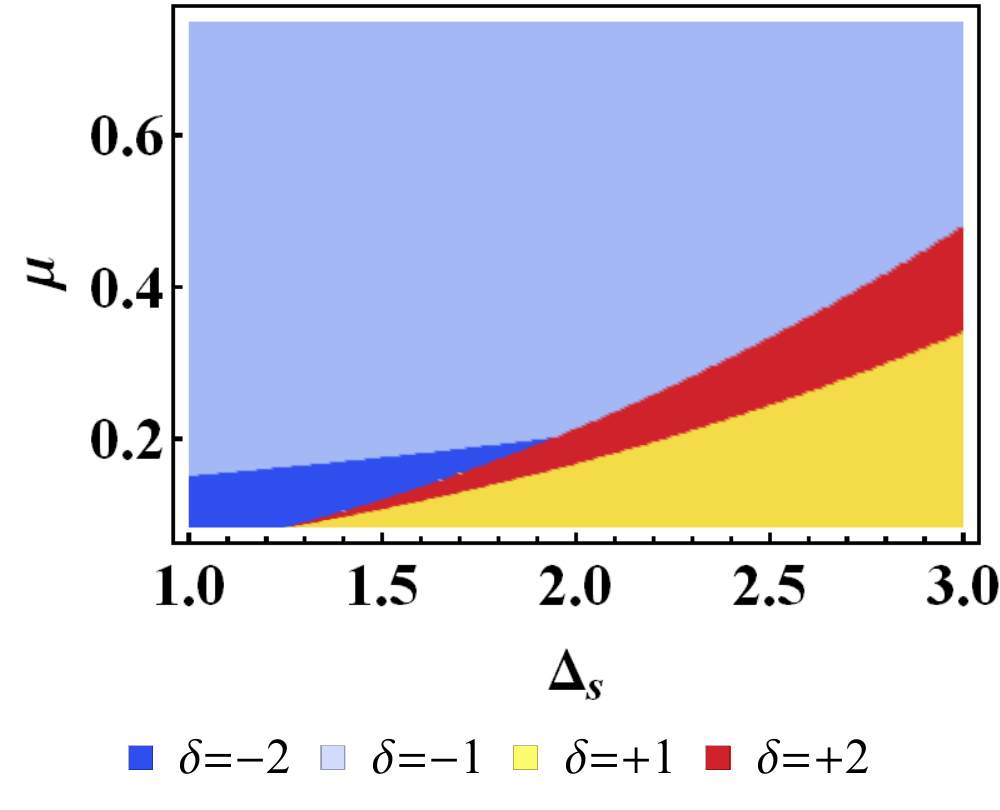}		
	\caption{(Color online) The phase diagram of a 1D metallic NW deposited on top of a 2D SC substrate. We plot the value of the topological index $\delta$ defined in Eq.~(9) of the main text as a function of  $\Delta_s$, and of the chemical potential $\mu$, which is taken to be the same for the NW and the substrate. We note that all the four possible phases corresponding to $\delta = \pm 1, \pm 2$ can form due to the Fermi momenta mismatch occurring because of different masses of quasiparticles in the SC and the NW. We have set $M=1.5, m=9, \varkappa=2.5$, and $t_{\mathrm{tun}}=0.5$.}
	\label{figureA2}
\end{figure}

In the main text we have proposed to obtain a Fermi momentum mismatch between $p_{\sigma_0}^{\mathrm{ind}}$ and $p_F$ by fixing the chemical potential of the substrate while tuning the chemical potential of  the NW, however this can be achieved also if the NW and the substrate have the same chemical potential but different band structures (e.g. different masses), and the difference between the Fermi momenta is achieved by tuning the common chemical potential; this is easier from an experimental perspective. An example of a corresponding phase diagram is plotted in Fig.~\ref{figureA2}, note that indeed all the four combinations of topological phases are recovered. 

\section{Non-TRS $p$-wave} \label{AppendixNonTRSpwave}

\begin{figure}[h!]
	\centering		\includegraphics[width=0.4\columnwidth]{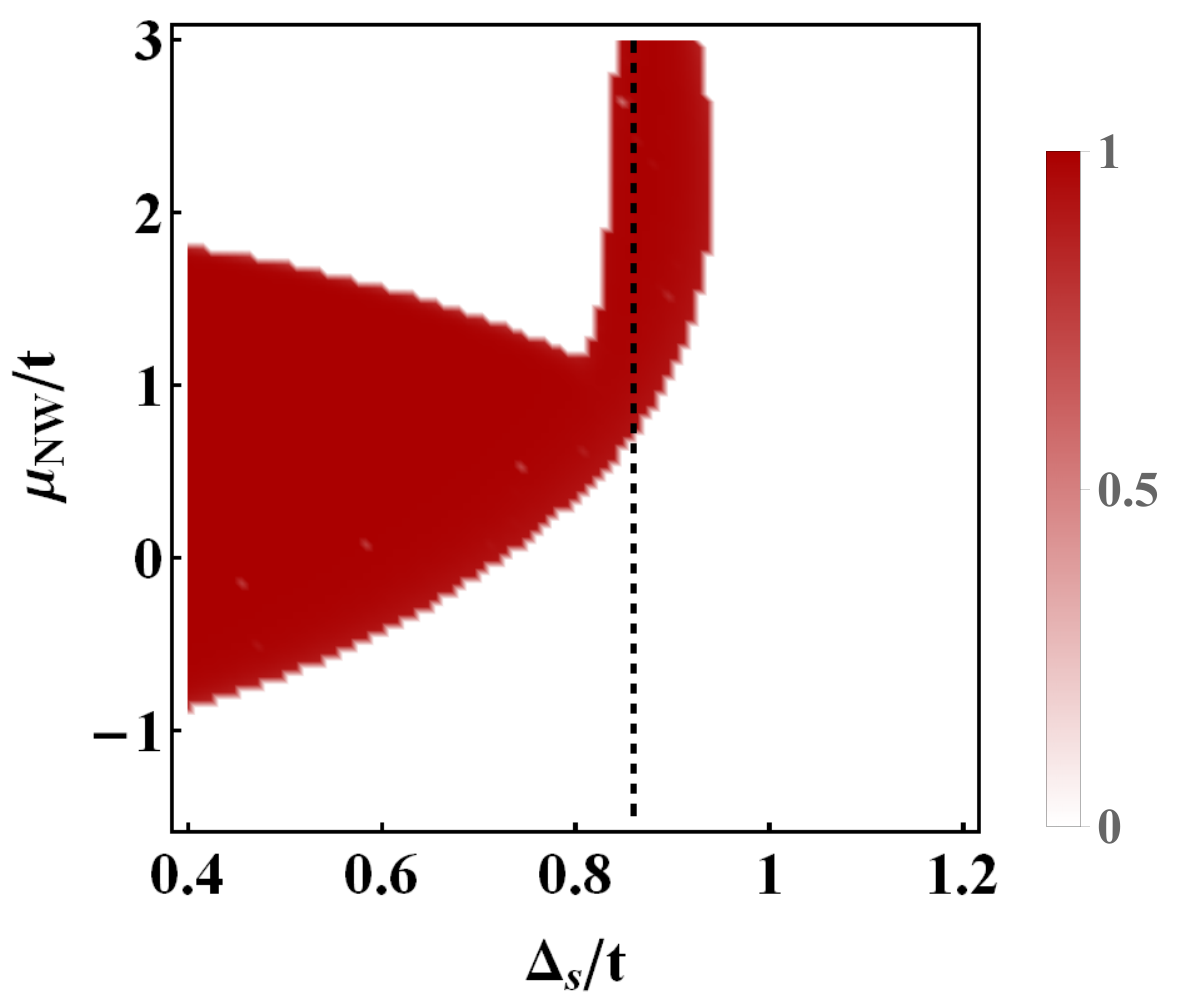}	
	\caption{The phase diagram of a 1D metallic NW deposited on top of a 2D SC substrate with a mixed $s$-wave and TRS-breaking $p$-wave pairing. Axes and color codes are the same as in Fig.~\ref{phdSO}. 
	 The black dashed line at $\Delta_s \sim 0.86t$ marks the  bulk topological phase transition.  We set $\mu_{\mathrm{SC}}=3, \Delta_t = 0.5$  and $t_{\mathrm{tun}}=2$.}
	\label{phdperp}
\end{figure}

As we have already discussed in the main text, not only time-reversal-symmetric (TRS) $s+p$--mixtures in the substrate may induce Majorana pairs at the ends of a proximitized NW. Below we consider a metallic nanowire on top of a substrate with a TRS-breaking $s+p$--wave pairing described by the following Hamiltonian:
\begin{eqnarray}
\nonumber \mathcal{H}_{\mathrm{TB}} = \sum\limits_{\bs{r}\in SC,\sigma} -\mu_{\mathrm{SC}} \Psi^\dag_{\bs{r}\sigma} \Psi_{\bs{r}\sigma} + \sigma \Delta_s \Psi_{\bs{r}\sigma} \Psi_{\bs{r},-\sigma} - t\sum\limits_{\bs{r}\in SC,\sigma}\left( \Psi^\dag_{\bs{r}\sigma} \Psi_{\bs{r}+\bs{x},\sigma} +\Psi^\dag_{\bs{r}\sigma} \Psi_{\bs{r}+\bs{y},\sigma} \right) \phantom{aaaaaaaaaaaaaaaaaaaaaaaaa} \\
-\Delta_t \negthickspace \sum\limits_{\bs{r}\in SC,\sigma} \left( i\Psi_{\bs{r}\sigma} \Psi_{\bs{r}+\bs{x},-\sigma} + \Psi_{\bs{r}\sigma} \Psi_{\bs{r}+\bs{y},-\sigma} \right)  + \negthickspace \sum\limits_{x\in NW,\sigma}\negthickspace -\mu_{\mathrm{NW}} c^\dag_{x\sigma} c_{x\sigma} - tc^\dag_{x\sigma} c_{x+1,\sigma} + t_{\mathrm{tun}} \negthickspace \sum\limits_{x \in NW,\sigma} \Psi^\dag_{x, y=0,\sigma} c_{x,\sigma} + \mathrm{H.c.} \phantom{aaa}
\label{TBHamnonTRS}
\end{eqnarray}
We demonstrate in Fig.~\ref{phdperp} that even when the time reversal symmetry is broken in the substrate in the presence of chiral $p$-wave pairing, Majorana pairs still form at the ends of the wire, and a phase diagram similar to that presented in Fig.~\ref{phdlattice} of the main text can be constructed. For a non-TRS $p$-wave SC these pairs cannot anymore be referred to as Kramers pairs, since it is no longer the TRS that protects them. However, they are protected by a different symmetry denoted 'magnetic mirror symmetry' (a weaker crystalline symmetry). For details we refer the reader to Refs.~[\onlinecite{Bernevig2014,Li2014,Andolina2017}].

\section{Derivation of the effective Hamiltonian for the nanowire obtained by integrating out the $s+p$-wave substrate} \label{AppendixDerivation}
\subsection{Model}
We consider a 1D nanowire (NW) with Rashba spin-orbit coupling on top of an $s+p$-wave 2D superconductor (SC). Writing the modified Bogoliubov-de Gennes Hamiltonian for the SC in the Nambu basis $\left\{ \Psi_{\bs p \uparrow},\Psi_{\bs p\downarrow}, \Psi^\dag_{-\bs p\downarrow}, -\Psi^\dag_{-\bs p \uparrow} \right\}^\mathrm{T}$ we get:
\begin{align*}
\mathcal{H}_{\mathrm{SC}} = \bpm \xi_p & 0 & \Delta_s & i\varkappa p_- \\ 
		 0 & \xi_p & -i\varkappa p_+ & \Delta_s \\
		 \Delta_s & i\varkappa p_- & -\xi_p & 0 \\
		 -i\varkappa p_+ & \Delta_s & 0 & -\xi_p 
	\epm,
	\quad \xi_p = \frac{p^2}{2M}-\frac{p_F^2}{2M}
\end{align*}
where $p_F$ is the Fermi momentum, $p_\pm = p_x \pm i p_y$, $\Psi_{\bs p \sigma}$ creates an electron with momentum $\bs p$ and spin $\sigma = \uparrow,\downarrow$. The singlet order parameter is denoted $\Delta_s$, whereas the triplet pairing is denoted $\varkappa$. We write the Hamiltonian of the NW in a similar Nambu basis $\left\{ c_{p_x \uparrow},c_{p_x \downarrow}, c^\dag_{-p_x \downarrow}, -c^\dag_{-p_x \uparrow} \right\}^\mathrm{T}$:
\begin{align*}
\mathcal{H}_{\mathrm{NW}} = 
	\bpm 
		\eta_{p_x} & 0 & 0 & 0 \\ 
		 0 & \eta_{p_x} & 0 & 0 \\
		 0 & 0 & -\eta_{p_x} & 0 \\
		 0 & 0 & 0 & -\eta_{p_x} 
	\epm,
	\quad \eta_{p_x} = \frac{p_x^2}{2m}-\mu
\end{align*}
where $\mu$ is the chemical potential.
We assume that tunnelling between the NW and the SC is allowed, and that the momentum along $x$ axis is conserved. The corresponding Hamiltonian is given by
\begin{align*}
\mathcal{H}_\mathrm{T} = t_{\mathrm{tun}}\sum\limits_\sigma \int \frac{d\bs p}{(2\pi)^2} \left[ \Psi^\dag_{\bs p \sigma} c_{p_x\sigma} + c^\dag_{p_x\sigma} \Psi_{\bs p \sigma} \right]
\end{align*}

In what follows we study how the interplay between different types of pairing in the SC affects the induced superconductivity in the NW.

\subsection{Path integral approach}

We start by writing the partition function in terms of the path integral:
\begin{align}
\mathcal{Z} = \int \mathcal{D} c^\dag_{p_x}\mathcal{D} c_{p_x} \int \mathcal{D} \Psi^\dag_{\bs p}\mathcal{D} \Psi_{\bs p} \,e^{i \left( \mathcal{S}_{\mathrm{NW}}^0 + \mathcal{S}_{\mathrm{SC}}^0 + \mathcal{S}_{\mathrm{T}} \right)},
\label{partitionfunction}
\end{align}
where $\mathcal{S}_{\mathrm{NW}}^0, \mathcal{S}_{\mathrm{SC}}^0$ and $\mathcal{S}_{\mathrm{T}}$ are the actions for the NW, SC and respectively for the NW-SC tunneling: 
\begin{align*}
\mathcal{S}_{\mathrm{SC}}^0 &= \frac{1}{2} \int \frac{d\bs p}{(2\pi)^2} \frac{d\omega}{2\pi} \Psi_{\bs p}^\dag \left[ G_{\mathrm{SC}}^0 \right]^{-1} \Psi_{\bs p} \\
\mathcal{S}_{\mathrm{NW}}^0 &= \frac{1}{2} \int \frac{d p_x}{2\pi} \frac{d\omega}{2\pi} c_{p_x}^\dag \left[ G_{\mathrm{NW}}^0 \right]^{-1} c_{p_x} \\
\mathcal{S}_{\mathrm{T}} &= \frac{1}{2} \int \frac{d\bs p}{(2\pi)^2} \frac{d\omega}{2\pi} \left[  \Psi_{\bs p}^\dag \mathcal{H}_\mathrm{T} c_{p_x} +  c_{p_x}^\dag \mathcal{H}_\mathrm{T}^\dag \Psi_{\bs p} \right].
\end{align*}
We use a notation similar to the previous subsection, however here we keep in mind that $c_{p_x} \equiv \left( c_{p_x \uparrow},c_{p_x \downarrow}, c^\dag_{-p_x \downarrow}, -c^\dag_{-p_x \uparrow}\right)^\mathrm{T}$ and $\Psi_{\bs p} \equiv \left( \Psi_{\bs p \uparrow},\Psi_{\bs p\downarrow}, \Psi^\dag_{-\bs p\downarrow}, -\Psi^\dag_{-\bs p \uparrow} \right)^\mathrm{T}$.
The bare Green's functions in energy-momentum space can be written as
$$
G_{\mathrm{SC}}^0(\omega, \bs p) = \left[\omega + i0 - \mathcal{H}_{\mathrm{SC}}(\bs p) \right]^{-1}, \quad G_{\mathrm{NW}}^0(\omega, p_x) = \left[\omega + i0 - \mathcal{H}_{\mathrm{NW}}(p_x) \right]^{-1}.
$$
Hereinafter we integrate out the superconducting degrees of freedom and we obtain an effective action for the NW, as well as its effective Green's function.

First, we rewrite the path integral in Eq.~(\ref{partitionfunction}) as
\begin{align*}
\mathcal{Z} = \int \mathcal{D} c^\dag_{p_x}\mathcal{D} c_{p_x} e^{i \mathcal{S}_{\mathrm{NW}}^0} \int \mathcal{D} \Psi^\dag_{\bs p}\mathcal{D} \Psi_{\bs p} \,e^{i \left(\mathcal{S}_{\mathrm{SC}}^0 + \mathcal{S}_{\mathrm{T}} \right)},
\end{align*}
and then we perform the inner integration over the Grassmann variables $\Psi_{\bs p}^\dag$ and $\Psi_{\bs p}$:
\begin{align*}
&\int \negthickspace \mathcal{D} \Psi^\dag_{\bs p}\mathcal{D} \Psi_{\bs p} \,e^{i \left(\mathcal{S}_{\mathrm{SC}} + \mathcal{S}_{\mathrm{T}} \right)} \equiv \int \negthickspace \mathcal{D} \Psi^\dag_{\bs p}\mathcal{D} \Psi_{\bs p} \,\exp \left\{ \frac{1}{2} \int\negthickspace \frac{d\bs p}{(2\pi)^2} \frac{d\omega}{2\pi} \Psi_{\bs p}^\dag \left[ G_{\mathrm{SC}}^0 \right]^{-1} \Psi_{\bs p}  + \left[  \Psi_{\bs p}^\dag \mathcal{H}_\mathrm{T} c_{p_x} +  c_{p_x}^\dag \mathcal{H}_\mathrm{T}^\dag \Psi_{\bs p} \right] \right\}
\end{align*}
It is known that for a given complex-valued matrix $A$ and Grassmann vectors $\zeta$ and $\xi^\dag$ the integral
$$
\int\negthickspace \mathcal{D}\eta^\dag \mathcal{D}\eta \exp \left[ \eta^\dag A \eta + \eta^\dag \zeta + \xi^\dag \eta \right] = \det A\, \cdot \exp \left[ -\xi^\dag A^{-1} \zeta \right].
$$
To obtain a similar expression we denote $A \equiv \frac{i}{2} \left[ G_{\mathrm{SC}}^0 \right]^{-1} $ and $\zeta = \frac{i}{2} \mathcal{H}_\mathrm{T} c_{p_{n_x}}, \xi^\dag =  \frac{i}{2}  c_{p_{n_x}}^\dag \mathcal{H}_\mathrm{T}^\dag $,  and we get:
\begin{align*}
\star &= \det \left[ \frac{i}{2} \left( G_{\mathrm{SC}}^0 \right)^{-1} \right] \cdot \exp \left\{\left( - \frac{i}{2}  c_{p_{n_x}}^\dag \mathcal{H}_\mathrm{T}^\dag\right) \left(\frac{2}{i} G_{\mathrm{SC}}^0  \right) \left(  \frac{i}{2} \mathcal{H}_\mathrm{T} c_{p_{n_x}} \right)\right\} = \\
& = \det \left[ \frac{i}{2} \left( G_{\mathrm{SC}}^0 \right)^{-1} \right] \cdot \exp \left\{ - \frac{i}{2}  c_{p_{n_x}}^\dag \mathcal{H}_\mathrm{T}^\dag G_{\mathrm{SC}}^0 \mathcal{H}_\mathrm{T} c_{p_{n_x}} \right\}
\end{align*}
Before inserting this expression back into the partition function, we employ the fact that the functional determinant is absorbed into the measure. Thus we have
\begin{align*}
\mathcal{Z} \propto \int \mathcal{D} c^\dag_{p_x}\mathcal{D} c_{p_x} e^{i \mathcal{S}_{\mathrm{NW}}^0} \cdot \exp \left\{ i \int\negthickspace \frac{dp_x}{2\pi} \frac{d\omega}{2\pi} c_{p_{n_x}}^\dag \left[ - \frac{1}{2} \int\negthickspace \frac{dp_y}{2\pi} \mathcal{H}_\mathrm{T}^\dag G_{\mathrm{SC}}^0 \mathcal{H}_\mathrm{T} \right] c_{p_{n_x}} \right\}
\end{align*}
Finally, the effective action and Green's function for the NW can be written as
\begin{align*}
\mathcal{S}_{\mathrm{NW}} = \mathcal{S}_{\mathrm{NW}}^0 + \int\negthickspace \frac{dp_x}{2\pi} \frac{d\omega}{2\pi} c_{p_{n_x}}^\dag \left[ - \frac{1}{2} \int\negthickspace \frac{dp_y}{2\pi} \mathcal{H}_\mathrm{T}^\dag G_{\mathrm{SC}}^0 \mathcal{H}_\mathrm{T} \right] c_{p_{n_x}}
\end{align*}
and
\begin{align*}
G_{\mathrm{NW}}^{-1} = \left( G^0_{\mathrm{NW}} \right)^{-1} - \int\negthickspace \frac{dp_y}{2\pi} \mathcal{H}_\mathrm{T}^\dag G_{\mathrm{SC}}^0 \mathcal{H}_\mathrm{T} 
\end{align*}

\subsection{Effective Green's function for the nanowire}

As we have just demonstrated above, within the path integral formalism, integrating out the superconducting degrees of freedom is equivalent to integrating the retarded Green's function of the superconductor over $p_y$ -- the momentum along the direction perpendicular to the axis of the NW. Therefore, we can write the "perturbed" retarded Green's function of the NW in the following form:
\begin{align*}
G_{\mathrm{NW}}(\omega, p_x) = \left[\omega + i0 - \mathcal{H}_{\mathrm{NW}}(p_x) - t_{\mathrm{tun}}^2 \sum\limits_{p_y} G_{\mathrm{SC}}^0(\omega, p_x, p_y) \right]^{-1},
\end{align*}
where we defined
\begin{align*}
G_{\mathrm{SC}}^0(\omega, p_x, p_y) = \left[\omega + i0 - \mathcal{H}_{\mathrm{SC}}(p_x, p_y) \right]^{-1}.
\end{align*}\\

We start by rewriting the retarded Green's function of the SC as $G_{\mathrm{SC}}^0 = \frac{1}{2} \sum\limits_{\sigma = \pm} G_0^{\sigma}$, where
\begin{align*}
G_0^{\sigma}(\omega,\bs p) = -\frac{1}{\xi_p^2 + (\Delta_s + \sigma \varkappa p)^2 - \omega^2}
\begin{pmatrix}
	1 & i \sigma e^{-i\phi_{\bs p}} \\
	-i \sigma e^{i\phi_{\bs p}} & 1
\end{pmatrix}
\otimes
\begin{pmatrix}
	\omega + \xi_p & \Delta_s + \sigma \varkappa p \\
	\Delta_s + \sigma \varkappa p & \omega - \xi_p
\end{pmatrix},
\end{align*}
with $p \equiv \sqrt{p_y^2 + p_x^2}$ and $e^{i s \phi_{\bs p}} \equiv \frac{p_x + i s p_y}{p}$ ($s = \pm 1$). To integrate this function over $p_y$ we linearize $\xi_p$ around the Fermi energy $\xi_p = v_F (p - p_F)$, where $v_F \equiv p_F/m$ is the Fermi velocity. There are four different integrations to perform, namely
\begin{align*}
X_0^\sigma &=  -\int \frac{dp_y}{2\pi} \frac{1}{\xi_p^2 + (\Delta_s + \sigma \varkappa p)^2 - \omega^2} \\
X_1^\sigma &=  -\int \frac{dp_y}{2\pi} \frac{\xi_p}{\xi_p^2 + (\Delta_s + \sigma \varkappa p)^2 - \omega^2} \\
X_2^\sigma &=  -\int \frac{dp_y}{2\pi} \frac{e^{i s \phi_{\bs p}}}{\xi_p^2 + (\Delta_s + \sigma \varkappa p)^2 - \omega^2}\\
X_3^\sigma &=  -\int \frac{dp_y}{2\pi} \frac{\xi_p\,e^{i s \phi_{\bs p}}}{\xi_p^2 + (\Delta_s + \sigma \varkappa p)^2 - \omega^2}
\end{align*}

We start by considering the case of $p_x \neq 0$ and we modify the integrand
$$
-\int \frac{dp_y}{2\pi} \frac{(\bullet)}{\xi_p^2 + (\Delta_s + \sigma \varkappa p)^2 - \omega^2} = \Bigg| p_y = p_x \sh q\Bigg| = -\frac{1}{v_F^2 + \varkappa^2} \frac{1}{|p_x|} \int \frac{dq}{2\pi} \frac{\ch q}{(\ch q + A_\sigma)^2+\Omega_\sigma^2} (\bullet),
$$
where
$$
A_\sigma \equiv \frac{\sigma \varkappa \Delta_s - p_F v_F^2}{(v_F^2 + \varkappa^2)|p_x|}, \quad \Omega_\sigma \equiv \frac{\sqrt{\left(\Delta_{\mathrm{eff}}^\sigma\right)^2 - \omega^2}}{|p_x|\sqrt{v_F^2 + \varkappa^2}}, \quad \Delta_{\mathrm{eff}}^\sigma \equiv \frac{\left|\Delta_s + \sigma \varkappa p_F \right|}{\sqrt{1+\varkappa^2/v_F^2}}
$$
$$
(\bullet) = 1, \quad (\bullet) = v_F(|p_x|\ch q - p_F), \quad (\bullet) = \frac{p_x}{|p_x|} \frac{1+is \sh q}{\ch q}, \quad (\bullet) = v_F(|p_x|\ch q - p_F)\frac{p_x}{|p_x|} \frac{1+is \sh q}{\ch q}
$$
for $X_0, X_1, X_2$ and $X_3$ correspondingly. For the sake of simplicity, in what follows we consider only in-gap energies, i.e. $|\omega| < \Delta_{\mathrm{eff}}^\sigma $, and we compute the zeroth integral
\begin{align*}
X_0^\sigma = \frac{1}{v_F^2 + \varkappa^2} \frac{1}{|p_x|} \frac{2}{\pi \Omega_\sigma} \im \left[ \frac{A_\sigma + i\Omega_\sigma}{\sqrt{1-(A_\sigma + i\Omega_\sigma)^2}}\arctg \frac{-1 + A_\sigma + i\Omega_\sigma}{\sqrt{1-(A_\sigma + i\Omega_\sigma)^2}}\right].
\end{align*}
We proceed with the 1st integral and we note that
$$
X_1^\sigma = -p_F v_F X_0^\sigma - \frac{v_F}{v_F^2 + \varkappa^2} \int \frac{dq}{2\pi} \frac{\ch^2 q}{(\ch q + A_\sigma)^2+\Omega_\sigma^2},
$$
where the last integral has a logarithmic UV divergence. To calculate it we introduce a cutoff in the following way
$$
\int \frac{dq}{2\pi} \frac{\ch^2 q}{(\ch q + A_\sigma)^2+\Omega_\sigma^2} = \lim\limits_{\Lambda \to +\infty} \int \frac{dq}{2\pi} \frac{\ch^2 q}{(\ch q + A_\sigma)^2+\Omega_\sigma^2} \cdot \frac{\Lambda^2}{(\ch q + A_\sigma)^2+\Lambda^2}.
$$
The integral can be done straightforwardly, and we obtain
\begin{align*}
\int \frac{dq}{2\pi} \frac{\ch^2 q}{(\ch q + A_\sigma)^2+\Omega_\sigma^2} \cdot \frac{\Lambda^2}{(\ch q + A_\sigma)^2+\Lambda^2} = \phantom{aaaaaaaaaaaaaaaaaaaaaaaaaaaaaaaaaaaaaaaa}\\
-\frac{2}{\pi} \frac{\Lambda^2}{\Lambda^2 - \Omega_\sigma^2} \left\{ \frac{1}{\Lambda} \im \left[  \frac{(A_\sigma + i\Lambda)^2}{\sqrt{1-(A_\sigma + i\Lambda)^2}}\arctg \frac{-1 + A_\sigma + i\Lambda}{\sqrt{1-(A_\sigma + i\Lambda)^2}} \right] - \right. \phantom{aaaaaaaaaaaaaaa}\\ 
\left. - \frac{1}{\Omega_\sigma} \im \left[  \frac{(A_\sigma + i\Omega_\sigma)^2}{\sqrt{1-(A_\sigma + i\Omega_\sigma)^2}}\arctg \frac{-1 + A_\sigma + i\Omega_\sigma}{\sqrt{1-(A_\sigma + i\Omega_\sigma)^2}} \right] \right\} = \star
\end{align*}
We expand the expression above keeping only the leading terms in $1/\Lambda$:
\begin{align*}
\star = \frac{2}{\pi \Omega_\sigma}  \im \left[  \frac{(A_\sigma + i\Omega_\sigma)^2}{\sqrt{1-(A_\sigma + i\Omega_\sigma)^2}}\arctg \frac{-1 + A_\sigma + i\Omega_\sigma}{\sqrt{1-(A_\sigma + i\Omega_\sigma)^2}} \right] + \frac{1}{\pi}\logn 2\Lambda + O\left(\frac{1}{\Lambda} \right)
\end{align*}
Along similar lines we compute the second integral
\begin{align*}
X_2^\sigma = -\frac{1}{v_F^2 + \varkappa^2} \frac{1}{p_x} \frac{2}{\pi \Omega_\sigma} \im \left[ \frac{1}{\sqrt{1-(A_\sigma + i\Omega_\sigma)^2}}\arctg \frac{-1 + A_\sigma + i\Omega_\sigma}{\sqrt{1-(A_\sigma + i\Omega_\sigma)^2}}\right].
\end{align*}
Note, that there is no dependence on the sign of $s$, even though it is present in the definition. The third integral can be expressed in terms of the zeroth and the second ones:
\begin{align*}
X_3^\sigma = v_F (p_x X_0^\sigma - p_F X_2^\sigma)
\end{align*}
To obtain the integrals computed above in the limit of $p_x = 0$, we take the limit $\lim\limits_{p_x \to 0}$ (it is not difficult to see that those limits exist). Finally, we write the retarded Green's function integrated over $p_y$ in terms of all the integrals computed above:
\begin{align}
\nonumber &\int \frac{dp_y}{2\pi} G_0^{\sigma}(\omega,\bs p) = \\ \nonumber
\nonumber &\phantom{a} \\
\nonumber &= 
\begin{pmatrix}
	\omega X_0^\sigma + X_1^\sigma & i\sigma (\omega X_2^\sigma + X_3^\sigma) & \Delta_\sigma X_0^\sigma + \sigma \tilde{\varkappa} X_1^\sigma & i\sigma \left(\Delta_\sigma X_2^\sigma + \sigma \tilde{\varkappa} X_3^\sigma \right) \\
	-i\sigma (\omega X_2^\sigma + X_3^\sigma) & \omega X_0^\sigma + X_1^\sigma & -i\sigma \left(\Delta_\sigma X_2^\sigma + \sigma \tilde{\varkappa} X_3^\sigma \right) & \Delta_\sigma X_0^\sigma + \sigma \tilde{\varkappa} X_1^\sigma \\
		\Delta_\sigma X_0^\sigma + \sigma \tilde{\varkappa} X_1^\sigma & i\sigma \left(\Delta_\sigma X_2^\sigma + \sigma \tilde{\varkappa} X_3^\sigma \right) & \omega X_0^\sigma - X_1^\sigma & i\sigma (\omega X_2^\sigma - X_3^\sigma) \\
		-i\sigma \left(\Delta_\sigma X_2^\sigma + \sigma \tilde{\varkappa} X_3^\sigma \right) & \Delta_\sigma X_0^\sigma + \sigma \tilde{\varkappa} X_1^\sigma & -i\sigma (\omega X_2^\sigma - X_3^\sigma) & \omega X_0^\sigma - X_1^\sigma
\end{pmatrix} \equiv \\ 
\nonumber	&\phantom{a} \\ 
& \equiv X_0^\sigma \sigma_0 \otimes (\omega \tau_0 + \Delta_\sigma \tau_x) + X_1^\sigma \sigma_0 \otimes (\tau_z + \sigma \tilde{\varkappa} \tau_x) - \sigma X_2^\sigma \sigma_y \otimes  (\omega \tau_0 + \Delta_\sigma \tau_x) - X_3^\sigma \sigma_y \otimes (\sigma \tau_z + \tilde{\varkappa}\tau_x).
	\label{EffGf}	
\end{align}
where we denoted $\tilde{\varkappa} = \varkappa / v_F$ and $\Delta_\sigma \equiv \Delta_s + \sigma \varkappa p_F$ (not to confuse with $\Delta^\sigma_{\mathrm{eff}}$).\\

Bringing together all the expressions calculated above we have:
\begin{align*}
&X_0^\sigma = \frac{1}{v_F^2 + \varkappa^2}  \frac{2}{\pi \widetilde{\Omega_\sigma}} \im \left[ \frac{\widetilde{A_\sigma} + i\widetilde{\Omega_\sigma}}{\sqrt{p_x^2-(\widetilde{A_\sigma} + i\widetilde{\Omega_\sigma})^2}}\arctg \frac{-|p_x| + \widetilde{A_\sigma} + i\widetilde{\Omega_\sigma}}{\sqrt{p_x^2-(\widetilde{A_\sigma} + i\widetilde{\Omega_\sigma})^2}}\right] \\
\nonumber &\phantom{a} \\ 
&X_1^\sigma = -p_F v_F X_0^\sigma - \frac{v_F}{v_F^2 + \varkappa^2} \left\{ \frac{2}{\pi \widetilde{\Omega_\sigma}} \im  \left[ \frac{(\widetilde{A_\sigma} + i\widetilde{\Omega_\sigma} )^2}{\sqrt{p_x^2-(\widetilde{A_\sigma} + i\widetilde{\Omega_\sigma})^2}}\arctg \frac{-|p_x| + \widetilde{A_\sigma} + i\widetilde{\Omega_\sigma}}{\sqrt{p_x^2-(\widetilde{A_\sigma} + i\widetilde{\Omega_\sigma})^2}} \right]\negthickspace + \frac{1}{\pi} \logn 2\Lambda \right\} \\
\nonumber &\phantom{a} \\ 
&X_2^\sigma = -\frac{1}{v_F^2 + \varkappa^2} \frac{2}{\pi \widetilde{\Omega_\sigma}} \cdot p_x \cdot \im \left[ \frac{1}{\sqrt{p_x^2-(\widetilde{A_\sigma} + i\widetilde{\Omega_\sigma})^2}}\arctg \frac{-|p_x| + \widetilde{A_\sigma} + i\widetilde{\Omega_\sigma}}{\sqrt{p_x^2-(\widetilde{A_\sigma} + i\widetilde{\Omega_\sigma})^2}}\right] \\
\nonumber &\phantom{a} \\ 
&X_3^\sigma = v_F \left(p_x X_0^\sigma - p_F X_2^\sigma \right),
\end{align*}
where tildes mean that the corresponding functions are taken at $p_x = 1$, i.e. $\widetilde{A_\sigma} = A_\sigma(p_x=1)$ and $\widetilde{\Omega_\sigma} = \Omega_\sigma(p_x=1)$.

\subsection{Studying the induced pairing}

We now study the low-energy approximation for the NW. Following Ref.~[\onlinecite{Nakosai2013prl}] we assume that its effective Hamiltonian can be found as
$$
\mathcal{H}_{\mathrm{NW}}^{\mathrm{eff}} = -\left[ G_{\mathrm{NW}}(\omega = 0, p_x) \right]^{-1},
$$
where we keep all the terms up to those linear in $p_x$. The induced singlet pairing is given by  $\left(\mathcal{H}_{\mathrm{NW}}^{\mathrm{eff}}\right)_{13}$, whereas the induced triplet pairing is defined by $\left(\mathcal{H}_{\mathrm{NW}}^{\mathrm{eff}}\right)_{14}$. It is easy to check that the corresponding symmetries are respected, i.e. the singlet pairing term is even in $p_x$, whereas the triplet pairing term is odd.

We substitute all the functions in Eq.~(\ref{EffGf}) to obtain an effective low-energy Hamiltonian. The induced pairing terms can be found as
\begin{align*}
\Delta_s^{\mathrm{ind}} = \frac{t_{\mathrm{tun}}^2}{\pi v_F} \lim\limits_{p_x \to 0} \sum\limits_\sigma  \frac{1}{|\Delta_\sigma|} \im \left[ \frac{\Delta_\sigma - i\sigma \tilde{\varkappa} |\Delta_\sigma|}{1+\tilde{\varkappa}^2}\frac{\widetilde{A_\sigma} + i\widetilde{\Omega_\sigma}}{\sqrt{p_x^2-(\widetilde{A_\sigma} + i\widetilde{\Omega_\sigma})^2}}\arctg \frac{-|p_x| + \widetilde{A_\sigma} + i\widetilde{\Omega_\sigma}}{\sqrt{p_x^2-(\widetilde{A_\sigma} + i\widetilde{\Omega_\sigma})^2}}\right] \\
\varkappa^{\mathrm{ind}} = \frac{t_{\mathrm{tun}}^2}{\pi v_F} \lim\limits_{p_x \to 0} \sum\limits_\sigma  \frac{-\sigma}{|\Delta_\sigma|} \im \left[ \frac{\Delta_\sigma - i\sigma \tilde{\varkappa} |\Delta_\sigma|}{1+\tilde{\varkappa}^2} \frac{1}{\sqrt{p_x^2-(\widetilde{A_\sigma} + i\widetilde{\Omega_\sigma})^2}}\arctg \frac{-|p_x| + \widetilde{A_\sigma} + i\widetilde{\Omega_\sigma}}{\sqrt{p_x^2-(\widetilde{A_\sigma} + i\widetilde{\Omega_\sigma})^2}}\right] \\
\lambda^{\mathrm{ind}} = \frac{t_{\mathrm{tun}}^2}{\pi v_F} \lim\limits_{p_x \to 0} \sum\limits_\sigma  \frac{\sigma}{|\Delta_\sigma|} \re \left[\frac{|\Delta_\sigma| - i\sigma \tilde{\varkappa} \Delta_\sigma}{1+\tilde{\varkappa}^2} \frac{1}{\sqrt{p_x^2-(\widetilde{A_\sigma} + i\widetilde{\Omega_\sigma})^2}}\arctg \frac{-|p_x| + \widetilde{A_\sigma} + i\widetilde{\Omega_\sigma}}{\sqrt{p_x^2-(\widetilde{A_\sigma} + i\widetilde{\Omega_\sigma})^2}}\right] \\
\end{align*}
Note that the functions in square brackets are not analytical at $p_x=0$, and therefore we must keep the limit sign and we cannot just set $p_x=0$. Numerically it means that we evaluate these functions for a very small non-zero value of $p_x$. We define also a renormalized chemical potential
\begin{align*}
&\mu^{\mathrm{ind}} = \mu - \frac{t_{\mathrm{tun}}^2}{2} \lim\limits_{p_x \to 0} \sum\limits_\sigma X_1^\sigma = \\
&= \mu + \frac{t_{\mathrm{tun}}^2}{\pi v_F} \left\{ \frac{\logn 2\Lambda}{1+\tilde{\varkappa}^2} + \lim\limits_{p_x \to 0} \sum\limits_\sigma  \frac{1}{|\Delta_\sigma|} \re \left[\frac{|\Delta_\sigma| - i\sigma \tilde{\varkappa} \Delta_\sigma}{1+\tilde{\varkappa}^2} \frac{\widetilde{A_\sigma} + i\widetilde{\Omega_\sigma}}{\sqrt{p_x^2-(\widetilde{A_\sigma} + i\widetilde{\Omega_\sigma})^2}}\arctg \frac{-|p_x| + \widetilde{A_\sigma} + i\widetilde{\Omega_\sigma}}{\sqrt{p_x^2-(\widetilde{A_\sigma} + i\widetilde{\Omega_\sigma})^2}}\right]\right\}
\end{align*}
In terms of the constants defined above we write an effective low-energy Hamiltonian for the NW:
\begin{equation}
H_{\mathrm{NW}}^{\mathrm{eff}} = 
	\bpm \frac{p_x^2}{2m}-\mu^{\mathrm{ind}} & i\lambda^{\mathrm{ind}} p_x  & \Delta_s^{\mathrm{ind}} & i\varkappa^{\mathrm{ind}} p_x \\ 
		 -i\lambda^{\mathrm{ind}} p_x & \frac{p_x^2}{2m}-\mu^{\mathrm{ind}} & -i\varkappa^{\mathrm{ind}} p_x  & \Delta_s^{\mathrm{ind}} \\
		 \Delta_s^{\mathrm{ind}} & i\varkappa^{\mathrm{ind}} p_x  &  -\left( \frac{p_x^2}{2m}-\mu^{\mathrm{ind}} \right) & -i\lambda^{\mathrm{ind}} p_x \\
		 -i\varkappa^{\mathrm{ind}} p_x  & \Delta_s^{\mathrm{ind}} & i\lambda^{\mathrm{ind}} p_x & -\left( \frac{p_x^2}{2m}-\mu^{\mathrm{ind}} \right)
	\epm,
	\label{efflowenergyHNW}
\end{equation}
or equivalently
$$
H_{\mathrm{NW}}^{\mathrm{eff}} = \left( \frac{p_x^2}{2m}-\mu^{\mathrm{ind}} \right) \sigma_0 \otimes \tau_z - \lambda^{\mathrm{ind}} \, p_x \sigma_y \otimes \tau_z + \Delta_s^{\mathrm{ind}} \, \sigma_0 \otimes \tau_x - \varkappa^{\mathrm{ind}} \,  p_x \sigma_y \otimes \tau_x.
$$
To find the gap for the Hamiltonian in Eq.~(\ref{efflowenergyHNW}) we linearize its spectrum around two different Fermi momenta emerging due to non-zero induced spin-orbit coupling value:
$$
\Delta^{\mathrm{ind}}_{\mathrm{eff}} = \min\limits_{\sigma}\frac{ \left| \Delta^{\mathrm{ind}}_s - m \varkappa^{\mathrm{ind}} \left(\lambda^{\mathrm{ind}} - \sigma \sqrt{(\lambda^{\mathrm{ind}})^2 + 2\mu^{\mathrm{ind}}/m}\right) \right|}{\sqrt{1+\frac{(\varkappa^{\mathrm{ind}})^2}{(\lambda^{\mathrm{ind}})^2 + 2\mu^{\mathrm{ind}}/m}}}
$$

\end{document}